\documentclass[sigconf,manuscript,screen]{acmart}

\usepackage{multirow}  
\usepackage{graphicx}
\usepackage{booktabs}

\usepackage{subcaption} 

\AtBeginDocument{%
  }

\usepackage{graphicx} 
\usepackage{subcaption} 

\begin{document}

\title{GANPrompt: Enhancing Robustness in LLM-Based Recommendations with GANs-Enhanced Diversity Prompts}

\author{Xinyu Li}
\affiliation{%
  \institution{College of Management and Economics, Tianjin University}
  \city{Tianjin}
  \country{China}}
\email{lxyt7642@tju.edu.cn}

\author{Chuang Zhao}
\affiliation{%
  \institution{Department of Electronic and Computer Engineering, The Hong Kong University of Science and Technology}
  \city{Hong Kong}
  \country{China}}
\email{czhaobo@connect.ust.hk}

\author{Hongke Zhao}
\authornotemark[1]
\affiliation{%
  \institution{College of Management and Economics, Tianjin University}
  \city{Tianjin}
  \country{China}}
\email{hongke@tju.edu.cn}

\author{Likang Wu}
\affiliation{%
  \institution{College of Management and Economics, Tianjin University}
  \city{Tianjin}
  \country{China}}
\email{wulk@mail.ustc.edu.cn}

\author{Ming HE}
\affiliation{%
  \institution{AI Lab at Lenovo Research}
  \city{Beijing}
  \country{China}}
\email{heming01@foxmail.com}

\begin{abstract}
In recent years, Large Language Models (LLMs) have demonstrated remarkable proficiency in comprehending and generating natural language, with a growing prevalence in the domain of recommendation systems. 
However, LLMs still face a significant challenge called prompt sensitivity, which refers to that it is highly susceptible to the influence of prompt words. This inconsistency in response to minor alterations in prompt input may compromise the accuracy and resilience of recommendation models. To address this issue, this paper proposes \textbf{GANPrompt}, a multi-dimensional LLMs prompt diversity framework based on Generative Adversarial Networks (GANs). The framework enhances the model's adaptability and stability to diverse prompts by integrating GANs generation techniques with the deep semantic understanding capabilities of LLMs. GANPrompt first trains a generator capable of producing diverse prompts by analysing multidimensional user behavioural data. These diverse prompts are then used to train the LLMs to improve its performance in the face of unseen prompts. Furthermore, to ensure a high degree of diversity and relevance of the prompts, this study introduces a mathematical theory-based diversity constraint mechanism that optimises the generated prompts to ensure that they are not only superficially distinct, but also semantically cover a wide range of user intentions.
Through extensive experiments on multiple datasets, we demonstrate the effectiveness of the proposed framework, especially in improving the adaptability and robustness of recommendation systems in complex and dynamic environments. The experimental results demonstrate that GANPrompt yields substantial enhancements in accuracy and robustness relative to existing state-of-the-art methodologies.
\end{abstract}



\maketitle

\section{INTRODUCTION}

In recent years, Large language models (LLMs), such as OpenAI's GPT series \cite{brown2020language, ouyang2022training} and Meta's LLaMA series \cite{touvron2023llama,touvron2023llama,dubey2024llama}, have been extensively trained on vast textual datasets, enabling them to develop advanced capabilities in natural language understanding and generation \cite{zhao2023survey,chang2024survey,minaee2024large}. Studies have shown that these models not only possess a significant amount of world knowledge but also demonstrate remarkable generalization abilities \cite{hadi2023survey,xi2023rise}. When fine-tuned appropriately, LLMs can apply their embedded knowledge to address specific domain challenges and solve targeted problems\cite{ding2023parameter, liu2023pre}. As a result, LLMs have the potential to revolutionize recommendation systems \cite{wu2023survey}. Their advanced language comprehension allows them to generate recommendations that surpass the capabilities of traditional behavioral analysis. By leveraging rich semantic information, LLM-based recommendation systems can offer higher precision and more refined personalization \cite{zhang2023chatgpt}.

One of the most promising approaches to adapting LLMs for recommendation tasks is prompt-based fine-tuning \cite{bao2023tallrec}. Unlike traditional pre-training methods, which involve modifying all model parameters with large-scale textual data, prompt-based fine-tuning techniques, such as Low-Rank Adaptation (LoRA \cite{hu2021lora}), integrate task-specific modules into the original model architecture. This approach allows for targeted optimization of parameters based on domain-specific datasets. By adding these modules, the model can focus on refining particular aspects of its knowledge without the need for extensive computational resources. As a result, it enables the rapid acquisition of specialized domain expertise and allows for effective adaptation to the unique requirements of recommendation tasks \cite{zhang2023recommendation, lyu2023llm}.

Despite its efficiency, fine-tuning LLMs alone has inherent limitations \cite{ghosh2024closer, hadi2024large}. While it allows LLMs to leverage semantic knowledge for recommendation tasks, relying exclusively on semantic information often results in performance gaps when compared to traditional recommendation models \cite{vm2024fine,zhang2023collm}. Traditional models excel at predicting user preferences through collaborative filtering, which uncovers hidden relationships within user-item interactions—a critical factor for accurate recommendations \cite{zhou2018deep,tan2016improved,tang2018personalized,kang2018self,sun2019bert4rec}. LLMs, however, primarily operate in the semantic space and often struggle to model these collaborative relationships effectively, leading to suboptimal performance when fine-tuning is used alone for recommendations \cite{zhang2023collm}.

To address these limitations, researchers have proposed hybrid methods that combine LLMs with traditional collaborative models \cite{zhang2023collm}. These strategies aim to merge semantic and collaborative information by incorporating collaborative signals from traditional models into LLMs and then fine-tuning the LLMs to better understand these signals. By leveraging the strengths of both approaches, hybrid methods bridge the gap between semantic understanding and collaborative knowledge, enabling LLM-driven recommendation models to capture user preferences more effectively, thereby improving recommendation accuracy. The success of these hybrid methods relies on the integration of both semantic and collaborative information: semantic information is derived from carefully crafted, recommendation-specific prompt templates that guide LLMs towards the desired tasks, while collaborative signals provide LLMs with specialized knowledge of user interactions.

\begin{figure}[ht]
  \centering
  \includegraphics[width=1.0\linewidth]{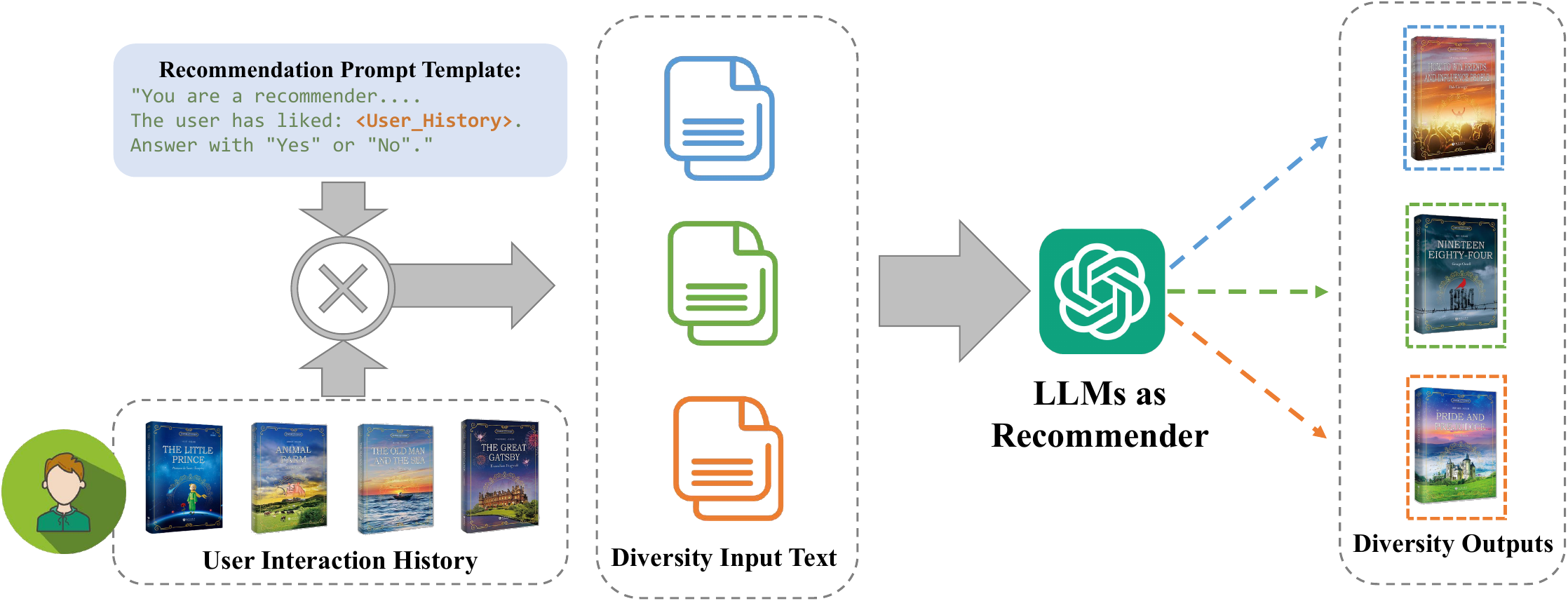}
  \caption{The different inputs gives different recommendation results in LLMs.}
  \label{fig1}
\end{figure}

However, significant challenges remain in effectively utilizing both semantic and collaborative information. One major issue is prompt sensitivity \cite{zhuo2024prosa}, where small changes in input prompts can lead to vastly different outputs, resulting in inconsistent, unpredictable, and potentially biased recommendations, as illustrated in Figure \ref{fig1}. Another challenge is the difficulty of efficiently integrating collaborative signals from traditional models. There is still a gap between the semantic space of LLMs and the latent collaborative space used by traditional models. Simple mapping layers often fail to fully integrate these collaborative signals, leading to suboptimal or even contradictory recommendation outcomes.

To address the issues outlined above, we propose a framework that combines \textbf{G}enerative \textbf{A}dversarial \textbf{N}etworks (GANs) with explicit collaborative guidance to enhance \textbf{Prompt}-based fine-tuning for LLM recommendation systems (\textbf{GANPrompt}). This model consists of two main components: GANs-based diversity encoder training and data generation, and fine-tuning LLMs with explicit collaborative guidance.

First, to address the issue of prompt sensitivity in LLM-based recommendation systems, we integrate a Generative Adversarial Network (GANs) framework with LLMs. This collaborative approach leverages both the semantic understanding of LLMs and the generative power of GANs to train a diversity-driven prompt generator. The generator is designed to produce a wide range of diverse and differentiated prompt samples, simulating the complex and varied user interactions encountered in real-world scenarios. These generated prompt samples are then used for fine-tuning the LLMs, improving the model's stability and accuracy when handling different, unseen prompts.

To ensure that the generated prompts are not only diverse but also relevant, the framework includes a mathematically-driven diversity constraint mechanism. This mechanism fine-tunes the prompts produced by the GANs using an optimization algorithm, ensuring that each prompt is distinct while also covering a broad spectrum of user intentions.

To address the gap in embedding spaces between LLMs and traditional collaborative model signals during the fine-tuning phase, we introduce expert-guided tokens into the optimization process of LLMs. These tokens are designed to inject knowledge from the collaborative model in a semantically aligned manner, enhancing the LLMs' understanding of the information provided by the traditional collaborative model.


Specifically, the contributions of this paper are as follows:
\begin{itemize}

\item This paper realises the deep integration of LLMs and GANs: by using the encoder of LLMs as a generator and constructing the corresponding discriminator to implement the generative adversarial network, we innovatively use the semantic parsing ability of LLMs and the generative characteristics of GANs to construct a multi-dimensional prompt generator. This combination not only enhances the diversity and quality of generated prompts, but also improves the adaptability and accuracy of the model in complex recommendation scenarios.

\item Proposing a distance-based diversity constraint: In order to ensure that the generated prompts are not only diverse but also semantically relevant, we propose a mathematical theory-based diversity constraint mechanism. This mechanism can effectively control the prompt generation process to ensure that each generated prompt covers a wide range of specific user intentions.

\item Explicitly Injecting the Performed Collaboration Knowledge: in order to bridge the semantic gap between LLMs and traditional collaboration models, we explicitly incorporate the recommendation results of traditional collaboration models into the prompt templates through the expert guidance token, which semantically guides the fine-tuning of the larger model.

\item Extensive experimental validation: we validate the effectiveness of the proposed framework through experiments conducted on multiple datasets. The experimental results show that GANPrompt outperforms existing state-of-the-art techniques in several recommendation tasks, especially in demonstrating greater robustness and higher accuracy when dealing with dynamic and diverse inputs.
\end{itemize}

\section{RELATED WORK}

In this paper, the related work involved includes the application of Fine-tuning LLMs in recommendation systems and generative adversarial networks.

\subsection{Fine-tuning Large Language Model as Recommendatiom System}

LLMs are transformer-based language models containing hundreds of billions (or more) parameters, which are self-supervised and trained on large amounts of textual data \cite{shanahan2024talking}, such as GPT-3 \cite{brown2020language}, PaLM \cite{chowdhery2023palm}, Galactica \cite{taylor2022galactica}, and LLaMA \cite{touvron2023llama}. Due to their absorption of vast amounts of textual knowledge, LLMs exhibit deep contextual semantic understanding and have demonstrated impressive performance in addressing complex tasks through text generation. Given the significant achievements of LLMs in the field of Natural Language Processing (NLP), their application in recommendation systems has become increasingly widespread \cite{ai2023information, dai2023uncovering, lin2023can, li2023large, wu2023survey,liu2023pre}.

Due to the high resource requirements for fully fine-tuning LLMs and the limited accessibility of closed-source models, researchers have explored equipping LLMs with sufficient contextual information through techniques such as instruction fine-tuning and in-context learning (ICL). These techniques enable LLMs to leverage their linguistic understanding, semantic prior knowledge, and generative capabilities to effectively address recommendation tasks. Wang et al. \cite{wang2022towards} propose a unified prompt learning paradigm for recommendation and conversational subtasks within a Conversational Recommendation System (CRS). By utilizing knowledge-enhanced prompts and conversation contexts, this approach provides the model with the necessary information to efficiently complete both tasks through prompt optimization. Geng et al. \cite{geng2022recommendation} introduce P5, a flexible and unified text-to-text sharing recommendation framework, using language as a medium. They construct multiple tasks into forms with the same language modeling objective and implement instruction-based recommendations via prompt learning. Gao et al. \cite{gao2023chat} innovatively convert user profiles and historical interactions into prompts, facilitating cross-domain transformation of user preferences expressed in textual form, thereby addressing the problem of data sparsity in cold-start scenarios. Zhang et al. \cite{zhang2023recommendation} take a user-centered approach to fine-tune LLMs by carefully designing instruction templates and generating fine-grained, user-personalized instruction data. Wang et al. \cite{wang2023recagent} explore the potential of human-like intelligence in LLMs, designing multiple behavioral prompts in real recommendation scenarios, enabling LLMs to simulate both the user and the recommender, evolving freely. Huang et al. \cite{huang2023recommender} provide multiple tools for building an interactive session recommendation system within LLMs. Wang et al. \cite{wang2023recmind} propose the Self-Inspiring algorithm, which enhances the planning ability of LLMs by allowing them to plan subsequent actions while considering all previously explored states. Yang et al. \cite{yang2023palr} utilize LLMs to summarize user preferences and initially retrieve a list of candidate items, which are then combined with other contexts for making recommendations.

While well-designed prompt templates for recommendation tasks can guide LLMs to learn the structure of recommendation tasks and perform well in zero- or few-shot scenarios \cite{lin2024rella}, simply exploiting semantic prior knowledge does not enable LLMs to achieve performance comparable to traditional recommendation models. Consequently, some researchers have adopted parameter-efficient fine-tuning, where recommendation knowledge is learned from the embedding space of LLMs by introducing soft prompts or fine-tuning modules.
Zhang et al. \cite{zhang2023prompt} provide an in-depth comparison of various prompt learning approaches in the news recommendation domain, showing that prompt templates with learnable virtual tokens achieve optimal performance due to their ability to adjust model parameters. Li et al. \cite{li2023personalized} present the first proposal to incorporate continuous user and item ID vectors as soft prompts into an input recommendation model, optimizing user and item representations in the potential embedding space of a language model. Li et al. \cite{li2023prompt} further distill the knowledge from discrete prompt templates of users and items into continuous vectors, effectively recognizing user and item names in LLMs, reducing inference time and computational cost while enhancing the performance of LLMs on recommendation tasks. Bao et al. \cite{bao2023tallrec} build upon ICL and introduce an additional parameter module (LoRA \cite{hu2021lora}) to achieve efficient fine-tuning of LLMs, enabling the rapid acquisition of recommendation capabilities with few samples and supporting multi-domain generalization. Xi et al. \cite{xi2024towards} transform and compress knowledge about user preferences and factual knowledge about items into augmentation vectors compatible with recommendation tasks, thereby enhancing the performance of subsequent recommendation models.

Although the approach discussed above demonstrates the feasibility of fine-tuning LLMs using recommendation data, it still falls short in certain aspects compared to traditional collaborative models. This limitation arises because LLMs continue to rely heavily on semantic prior knowledge and struggle to capture the collaborative information between users and items, which is crucial for recommendation systems.
Zhang et al. \cite{zhang2024agentcf} propose an agent-based collaborative learning method, where user and item agents are constructed, and the description texts of the agents are treated as model parameters. The forward and backward propagation processes are modeled through interactions and reflections between the agents. Li et al. \cite{li2023ctrl} leverage both tabular and textual data for fine-grained cross-modal knowledge alignment between traditional collaborative models and language models, integrating both collaborative and semantic signals for recommendation. Zhang et al. \cite{zhang2023collm} build upon the work of Bao et al. \cite{bao2023tallrec} and integrate collaboration information learned by traditional models into LLMs. This allows LLMs to synthesize both semantic priors and collaborative knowledge, thereby enhancing their performance in recommendation tasks and surpassing traditional models. Luo et al. \cite{luo2024integrating} apply data and prompt augmentation strategies to enrich both traditional models and LLMs. They employ an adaptive aggregation module to combine the predictions from both models, refining the final recommendation results and addressing the data sparsity and long-tail problems encountered by traditional collaborative models. Zhu et al. \cite{zhu2024collaborative} extend the vocabulary of LLMs by incorporating user and project ID tokens and applying mutual regularity pre-training based on soft+hard prompting strategies, thereby effectively capturing user/project collaboration and content information through language modeling. Luo et al. \cite{luo2023recranker} enhance the capability of LLMs for Top-k recommendations by sampling high-quality, representative, and diverse training examples for instruction-tuned LLMs, further augmenting the prompts with auxiliary information from traditional collaborative recommendation models.

\subsection{Generating Adversarial Networks}

Generative Adversarial Networks (GANs) are deep learning models grounded in zero-sum game theory \cite{goodfellow2014generative}. They combine two components—generators and discriminators—engaged in adversarial learning, where the optimal learning process can be framed as a minimax game problem \cite{li2017parallel}. The goal of this game is to find the Nash equilibrium \cite{nash1953two}, where both the generator and the discriminator reach an optimal point, thus improving model performance. The generator estimates the data sample distribution by learning from the adversarial feedback provided by the discriminator.

Xu et al. \cite{xu2017neural} introduces GANs for single-issue, short-text dialogues, where the generator's output is optimized through the design of an approximate latent layer, which helps generate responses with greater diversity. Xu \cite{xu2018diversity} further proposes DP-GANs, which differentiates between new and duplicate texts, assigning lower rewards to duplicate texts. This incentivizes the generation of more diverse and information-rich texts for tasks such as comment generation and dialogue generation.
Zhang et al. \cite{zhang2018generating} maximizes the variational information by introducing noise into the generator. This approach maximizes the Variational Information Maximization Objective (VIMO), which enhances the informativeness and diversity of the generated text.
Gurumurthy et al. \cite{gurumurthy2017deligan} introduces DeLiGANs, a model that improves the ability to model prior distributions by representing the latent space as a mixture of multiple Gaussian distributions. This enables the generator to sample from more complex latent distributions, thus generating more diverse samples, even with limited data.
Gu et al. \cite{gu2018dialogwae} constructs a GANs in the latent space of digital autoencoders, specifically Variational Autoencoders (VAEs), using GANs to model the data distribution. By combining Wasserstein distance and Gaussian mixture models (GMM), the approach achieves greater diversity and information richness in the generated responses.
Chen et al. \cite{chen2018adversarial} defines Feature-Mover's Distance (FMD) as a method for aligning the potential feature distributions of real and synthetic sentences. This method aims to overcome the training instability issue commonly associated with GANs, while improving diversity, as reflected in self-BLEU statistics \cite{zhu2018texygen}.

\section{METHOD}

In this section, we first formalise the problem and briefly introduce the overall framework, then go into the details of the submodules.

\subsection{Problem Definition and Overview of Framework}
\label{3.2}

\textbf{Problem Definition}. Let us consider a universe of users represented by the set $U$ and a collection of items denoted by the set $I$. Within this scenario, a particular user is indicated by $u$, where $u \in U$, and a specific item is denoted by $i$, where $i \in I$.  Each interaction between a user $u$ and an item $i$ is meticulously recorded in a tuple $(u, i, t)$, where $t$ represents the timestamp of the interaction. Additionally, each interaction record may also include a rating $r$ assigned by the user to the item, the rating $r$ is scaled from 1 to 5, i.e.,$r \in [1, 5]$. Each item also possesses an associated description file $d$, which encompasses either the title or a detailed descriptive text of the item. For each user $u$, there is a series of project interaction records $[i^u_1,i^u_2,i^u_3,...,i^u_n]$, which sorted by timestamp , while each interaction record also contains the corresponding user rating $r$ and item description $d$.

In this paper, we focus on the sequence recommendation task, which specifically refers to the use of information about a user's past history of interaction sequences to predict the next item that can be interacted with, as shown below:

$$P(i^u_{n+1}|[i^u_1,i^u_2,i^u_3,...,i^u_n],[d_1,d_2,d_3,...,d_n]),$$

For user $u$, $i^u_{n+1}$ is the predicted next target item and $P(\cdot)$ is the predicted interaction probability.

\begin{figure}[t]
  \centering
  \includegraphics[width=1.0\linewidth]{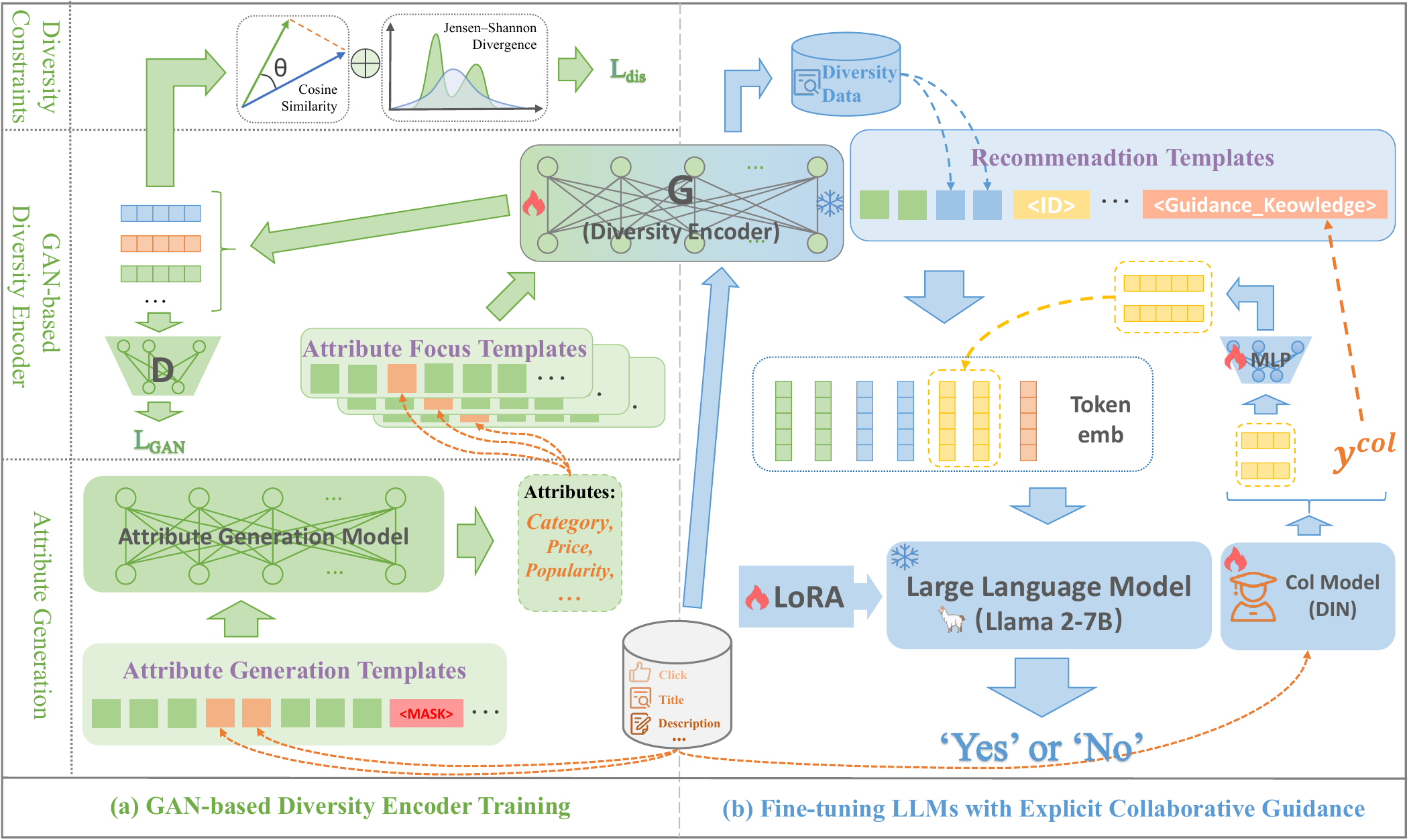}
  \caption{Overall framework diagram of the model.}
  \label{fig 2}
\end{figure}

\textbf{Overview of Framework}. This subsection presents the modelling framework and module details proposed in this paper in detail. Firstly, an overview of the framework proposed in this paper is given as Figure \ref{fig 2}. 
The framework proposed in this paper contains two major parts: \textit{GANs-based diversity encoder training} and \textit{LLMs-based recommendation fine-tuning with diversity data and collaborative knowledge}.
The construction of the \textit{GANs-based diversity encoder training} comprises three key components: attribute generation module, GANs-based encoder diversity module, and diversity constraint module. 
In the \textbf{recommendation task fine-tuning} phase, the trained diversity encoder is used to generate diversity data samples, which are then combined with collaborative information from the traditional recommendation model into a recommendation task prompt template to fine-tune the downstream LLMs-based recommendation model. The goal is to take full advantage of both diversity data and traditional model knowledge to collaboratively improve the robustness and accuracy of LLMs-based recommendations.

\subsection{GANs-based Diversity Encoder Training}
This section describes the training process of the GANs-based diversity encoder.

\subsubsection{Attribute Features Generation}


Employing LLMs to generate diverse attributes for data has been demonstrated to effectively enhance the diversity of the original dataset, thereby providing complex data support for fine-tuning downstream tasks\cite{gao2022self}. 
Consequently, the objective of this paper is to design augmented attribute prompts for various recommendation task datasets. By leveraging LLMs to generate these diverse attributes, we aim to bolster the diversity of the original data, which in turn enriches the complexity of the data for subsequent fine-tuning processes.

\begin{figure}[h]
  \centering
  \includegraphics[width=0.9\linewidth]{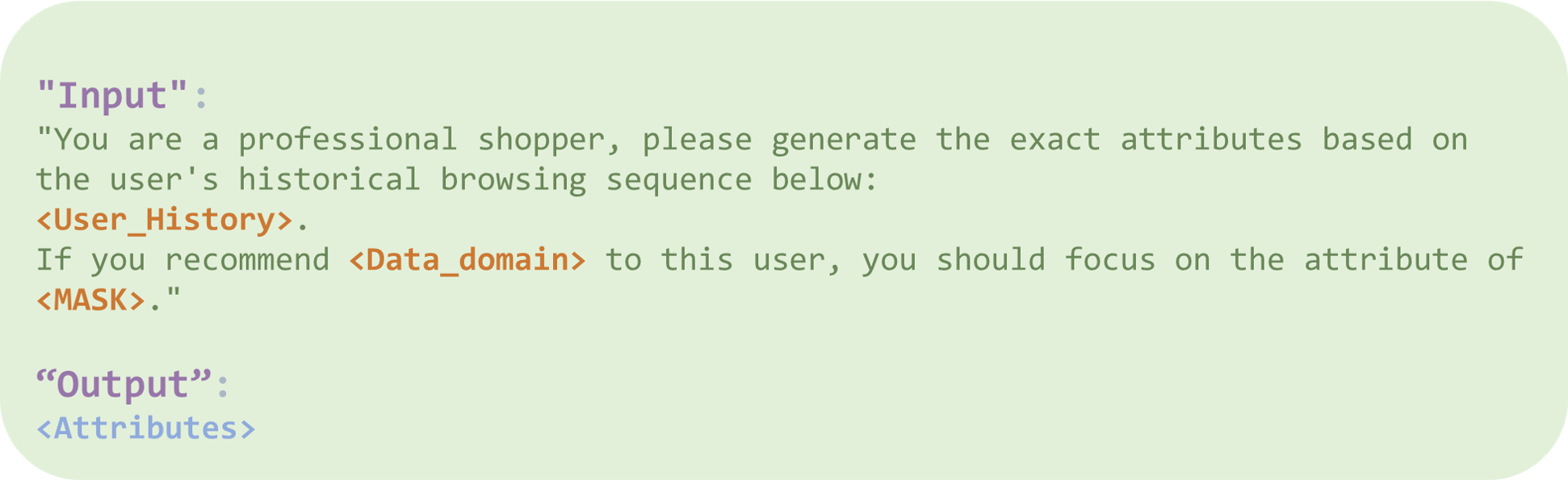}
  \caption{Attribute generate task Prompt Template for Sequential data.}
  \label{prompt 2}
\end{figure}

To be specific, in the context of sequence recommendation, the corresponding item description sequence $[d_1,d_2,d_3,.... .d_n]$ or a separate item description texts $d_i$ constructed from the original data and input into the following attribute generation template \ref{prompt 2}.

The $<User\_History>$ denotes the sequence of descriptions pertaining to the user's interaction history, while $<Data\_domain>$ signifies the target category of items present within the relevant recommendation dataset. In the context of attribute data generation, we employ a MASK prediction task framework. The $<MASK>$ token serves as a placeholder, which is utilized to forecast the attribute characteristics associated with the data sample. The $<Attributes>$ in the output represents the attributes that the model generates for the given data sample. It can be formalized as follow:

$$a_i^{seq} = LLMs_{Attr}(T_{Attr}^{seq} + [d_1^{u_i}, d_2^{u_i}, d_3^{u_i},...,d_n^{u_i}]),$$

where  $T_{Attr}^{seq}$ denotes the attribute generation template. The sequence $[d_1^{u_i}, d_2^{u_i}, d_3^{u_i}, \ldots, d_n^{u_i}]$ encapsulates the sequence data, which is assembled from item descriptions within the historical interaction sequence of user $u_i$. This sequence can be sourced from either item titles or descriptive texts. For the purposes of this paper, we have chosen to use item titles to construct our training data, imposing a maximum token limit of 20 per item title to maintain consistency. The $LLMs_{Attr}(\cdot)$ designates the LLMs utilized for attribute generation. In this study, we employ BERT~\cite{}, a pre-trained language model, as our $LLMs_{Attr}$. Furthermore, the symbol $a^{seq}$ represents the attribute text that is generated in correspondence with each sample within the sequence recommendation dataset.

Following the attribute generation process, we perform a manual curation to exclude any inappropriately generated attributes. Subsequently, we rank the generated attributes and select the attribute features that occur most frequently, specifically retaining those within the top-$k$ occurrences, to construct the training data for subsequent tasks. In this study, we set $k$ to 5.

\subsubsection{Diversity Encoder Training Based on GANs}
Although the method of attribute generation helps to increase the diversity of the data, there are still limitations in discrete attribute features. 
In this paper, inspired by Generative Adversarial Networks (GANs), we construct the encoder of the Large Language Model as a generator $ G_{LLMs_e}(\cdot)$, which is tasked with generating continuous vector representations for discrete textual data. Specifically, we create a multi-dimensional attribute dataset by combining original data with generated attribute features, subsequently producing continuous vector features for both the original data samples and their corresponding attribute data. To implement the zero-sum game process characteristic of GANs, we develop a corresponding discriminator $ D_{MLP}(\cdot)$ to differentiate between attribute samples of varying categories. Through the adversarial learning process, the generator's capability to enhance the diversity of data samples is further augmented. The detailed implementation process is as follows.

\textbf{Generation Process.}
Through the attribute features generation process, we can obtain $k+1$ datasets that include original data samples and $k$ attribute datasets. In the generation process, we take the data samples with added attributes as random noise and input them into the LLMs encoder to generate the latent features, which are in the following form for the sequence recommendation task, for example:

$$emb_{real}^{seq} = G_{LLMs_e}([d_1^{u_i}, d_2^{u_i}, d_3^{u_i},...,d_n^{u_i}]),$$

$$emb_{attr_n}^{seq} = G_{LLMs_e}(T_{GANs}^{seq} + a_n^{seq} + [d_1^{u_i}, d_2^{u_i}, d_3^{u_i},...,d_n^{u_i}]),$$

\begin{figure}[h]
  \centering
  \includegraphics[width=0.6\linewidth]{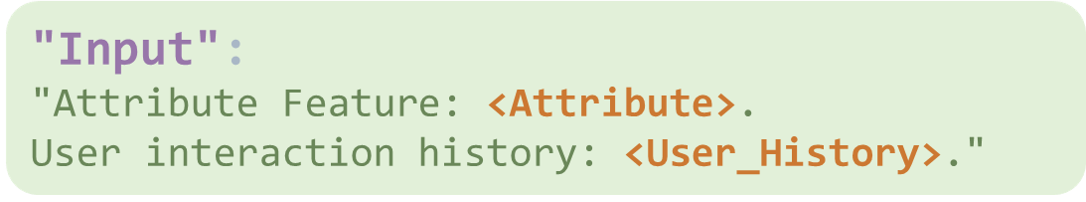}
  \caption{Attribute Feature data Builds Prompt Template for Sequential data.}
  \label{prompt 3}
\end{figure}

In this context, \( a_n^{seq} \) denotes the \( n \)-th attribute within the sequence recommendation task. Concurrently, \( T_{GANs}^{seq} \) is the adversarial learning prompt template constructed from attribute feature data, which is shown as Figure \ref{prompt 3}. This template facilitates the amalgamation of the user's interaction history with its respective attributes, where these attributes, once combined, are treated as noise. Ultimately, \( emb_{real}^{seq} \) signifies the potential feature vector generated for the interaction history of user \( u \), whereas \( emb_{attr_n}^{seq} \) represents the potential feature vector generated for the \( n \)-th attribute data associated with the user. The encoder used in this paper is the encoder of the T5 pre-trained language model\cite{raffel2020exploring}.

\textbf{Discriminative Process.}
Following the generation of potential feature vectors for the corresponding data, a simple discriminator was constructed to classify them for the adversarial learning zero-sum gaming process. The discriminator comprises three layers of multilayer perceptrons (MLPs):

$$D_{MLP}= \sigma_3(W_3 \sigma_2(W_2 \sigma_1(W_1x+b_1)+b_2)+b_3),$$

The input vector, $x$, is fed into the network, where it is processed by a series of layers. Each layer is represented by a weight matrix, bias vector and activation function, which are denoted by $w_i$, $b_i$ and $\sigma$, respectively. The index $i$ represents the layer number, which ranges from 1 to 3. Subsequently, the output vector of the generator is fed into the discriminator, which then outputs the predicted category.

$$y_{GANs} = D_{MLP}(emb^{seq}),$$

The input of the data feature vector to the discriminator, denoted by $emb^{seq}$, is compared to the prediction result, $y_{GANs}$. For real data and data of the $k$ attribute, the discriminator is expected to recognize the input into the corresponding $k+1$ categories.

\subsubsection{Diversity Constraint}

Diversity is a fundamental problem in classification and clustering tasks, and its concept is strongly related to distance; as the distance between two samples increases, the similarity between them decreases, thus increasing the diversity between the samples\cite{xia2015learning}.
In order to make the diversity encoder more effective in expanding the differences between different samples, this paper introduces the cosine similarity distance and the JS dispersion from the perspective of mathematical theory to calculate the angle and information differences between different samples, in order to measure the diversity between samples. And it is used as the diversity constraint index in the encoder optimization process, so that the optimized diversity encoder can distinguish samples more effectively.

\textbf{Cosine similarity.}
Cosine similarity uses the cosine of the angle between two vectors to measure similarity. It is expected that there will be a small angle between two similar vectors $x_i^{\prime}$ and $x_i$, and the cosine similarity is defined as follows:

$$\cos (\theta)=\frac{\sum_{i=1}^{d} x_{i} \times x_{i}^{\prime}}{\sqrt{\sum_{i=1}^{d} x_{i}^{2}} \times \sqrt{\sum_{i=1}^{d} x_{i}^{\prime 2}}},$$

Where $d$ is the dimension between the vectors $x_i$ and $x_i^{\prime}$, $\theta$ is the angle between the vectors $x_i$ and $x_i^{\prime}$, and the similarity between the two vector samples can be measured using $cos(\theta)$. As $cos(\theta)$ decreases, the angle between the two vectors increases, which simultaneously implies an increase in the distance between the vectors and a weakening of the similarity, which in this paper will be viewed as an increase in the diversity between the vector samples.

However, basic cosine similarity has the serious problem of focusing only on the directions between patterns, which leads to the computation of the diversity metric between vectors being isolated. For this reason, this paper introduces Jensen-Shannon divergence to complement it.

\textbf{Jensen-Shannon Divergence.} 
The Jensen-Shannon divergence is a symmetric version of the KL (Kullback-Leibler Divergence) scatter, which is used to compute the loss of information between two probability distributions and thus measure the difference between them. For two similar vectors $x_i^{\prime}$ and $x_i$, the KL scatter uses $x_i^{\prime}$ to approximate $x_i$ and calculates how much information would be lost, thus the KL scatter is asymmetric, i.e., $D_{KL}(x_i^{\prime} | | x_i) \neq D_{KL}(x_i | | x_i^{ \prime})$. The specific formula is as follows:

$$D_{KL}(x_i \parallel x_i^{\prime}) = \sum x_i \log \frac{x_i}{x_i^{\prime}},$$

The Jensen-Shannon (JS) scatter measure is derived from the Kullback-Leibler (KL) scatter by introducing an intermediate distribution, which symmetrizes and normalizes the KL divergence. The specific formula is presented as follows: 

$$D_{JS}(x_i \parallel x_i^{\prime}) = \frac{1}{2} D_{KL}(x_i \parallel m) + \frac{1}{2} D_{KL}(x_i^{\prime} \parallel m),$$

$$m = \frac{1}{2}(x_i + x_i^{\prime}),$$

where $m$ is the mean distribution of the sample $x_i$ and $ x_i^{\prime}$.

\textbf{Weighted diversity constraints}.
Neither cosine similarity nor Jensen-Shannon divergence, when considered in isolation, can sufficiently capture the diversity among data vectors; therefore, the distance metric used in this paper combines cosine similarity and Jensen-Shannon divergence in the following form:

$$D_{total} = \alpha D_{cos}(x_i \parallel x_i^{\prime}) + \beta D_{JS}(x_i \parallel x_i^{\prime}),$$

where $\alpha$ and $\beta$ are the weights in the distance-metric synthesis process. By combining the cosine similarity and Jensen-Shannon divergence, this metric can synthesize the angular distance differences and information differences between sample vectors. This diversity metric is involved in the diversity encoder optimisation process for the construction of sample diversity in both the distance and the information dimensions.

\subsection{LLMs-based Recommendation Task Fine-tuning}
Most existing LLMs recommendation methods that utilize fine-tuning adhere to the natural language processing paradigm: Initially, a prompt template for the recommendation task is constructed to align the recommendation data with the input of the LLMs; subsequently, the prompt text is transformed into a sequence of tokens, and feature embeddings for both tokens and sentences are computed. Finally, the architecture of the LLMs undergoes fine-tuning through techniques such as LoRA to adapt it for downstream recommendation tasks. 

While fine-tuned LLMs can adeptly learn the paradigms of recommendation tasks and leverage corpus knowledge from textual data, they still struggle to comprehend collaborative information inherent in interactive datasets. This limitation results in strong performance during cold start scenarios but subpar accuracy in traditional recommendation settings. 

To address this issue, this paper draws on the work of CoLLM by employing a conventional recommendation model to derive embedded vectors representing user-project collaboration information. These vectors are then aligned with the input layer dimensions of the LLMs through a mapping layer, enabling the model to synergistically utilize both extensive text prediction knowledge and collaborative insights for enhanced recommendations. Consequently, it achieves superior accuracy in traditional recommendations while maintaining robust performance in cold start situations. 

Although the direct injection of collaborative signals can help supplement the LLMs with collaborative information, it is inevitable that the big model thinks in a way that deviates from the traditional model. In order to further better guide the LLMs to understand the traditional model's recommendation knowledge, we explicitly add the \textbf{guidance token} in the recommendation prompt template.

\begin{figure}[h]
  \centering
  \includegraphics[width=0.9\linewidth]{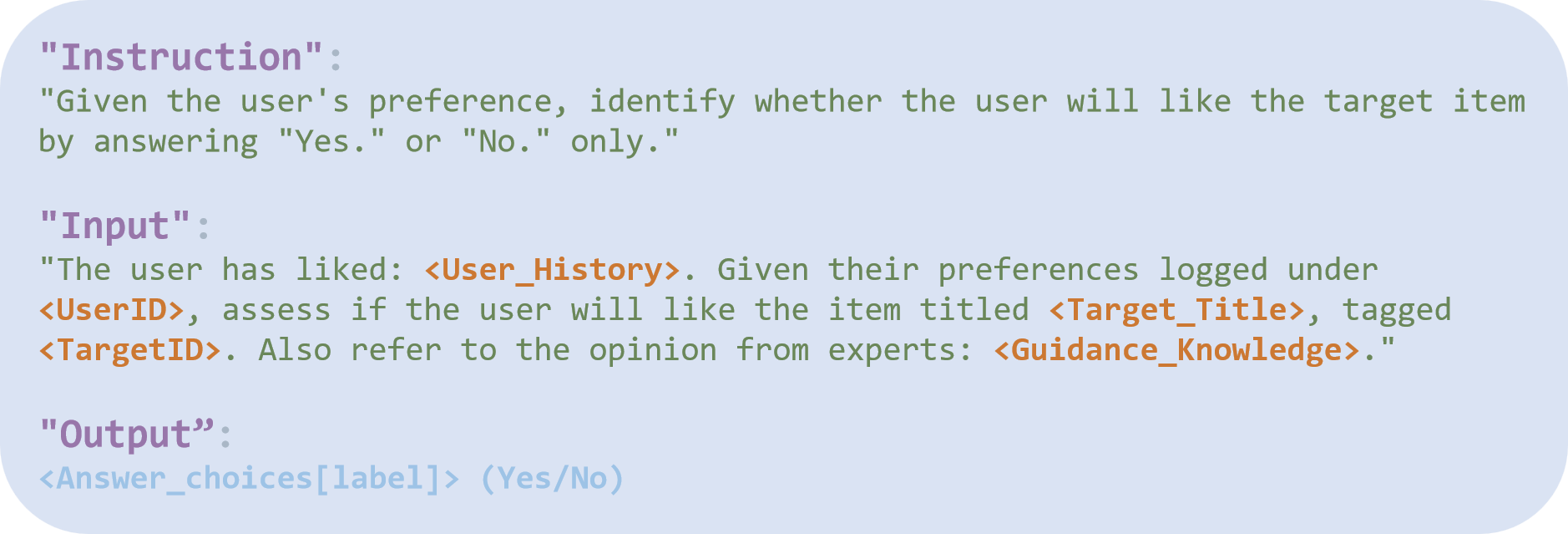}
  \caption{Sequential Recommendation Task Prompt Template .}
  \label{prompt 1}
\end{figure}

\textbf{Recommendation Prompt Template Definition.} 
Initially, we developed a prompt template for instruction fine-tuning to align LLMs with recommendation tasks, drawing on previous research \cite{wang2022towards,geng2022recommendation}. 
Specifically, we leverage the user's interaction sequence, denoted as $[i_1, i_2, i_3, \ldots, i_{n-1}]$, to forecast the subsequent interactive item, $i_n$.
The detailed structure of this prompt template is illustrated in Figure \ref{prompt 1}.

The $Instruction$ section serves as the directive for guiding the LLMs in performing recommendation tasks. 
The $Input$ section contains the primary information that the LLMs receives, where $<User\_History>$ represents the user's sequential interaction history, and $<Target\_Title>$ denotes the target item to be predicted. Both are constructed using the textual data of item titles and replace the corresponding placeholders in the template. $<UserID>$ and $<TargetID>$ are the identifiers for the user and the item, respectively. Here, we have drawn upon the work of~\cite{zhang2023collm} to incorporate collaborative interaction information between users and items into the corresponding ID tokens. This allows the LLMs to leverage collaborative information for more accurate recommendations. 
In addition, $<Guidance_Knowledge>$ denotes the knowledge received from the traditional model, and here we directly input the recommendation results from the traditional model into this token to explicitly guide the LLMs.
Finally, by accepting the logical probability output of the $(Yes/No)$ token from the $Output$ section of the LLMs, the prediction result for the target item is determined.

\textbf{Diversity data reconstruction.}
In Chapter \ref{3.2},  we employ an attribute generation method to produce a variety of attribute features for the original dataset. These enriched data are then utilized to train a diversity encoder that leverages generative adversarial networks and incorporates diversity constraints. During the recommendation task, we commence by reconstructing the original training dataset. For the specific case of the original project description $d_i$, we apply a diversity encoder with attribute knowledge diversity constraints to generate diverse representation vectors. Subsequently, we utilize the original decoder of the LLMs (T5) to decode these vectors, yielding a reconstructed project description $d_i^r$. The detailed process is delineated as follows: 

$$emb_{i}^{div} = G_{LLMs_e}(d_i),$$

$$d_{i}^{r} = G_{LLMs_d}(emb_{i}^{div}),$$
Here, \( emb_{i}^{div} \) represents the diversity feature vector generated for item \( i \) using the diversity encoder; \( G_{LLMs_d}(\cdot) \) denotes the decoder of the original LLMs, with the specific use of the T5 model's decoder in this context; \( d_{i}^{r} \) signifies the reconstructed project description text for item \( i \). Consequently, we integrate discrete attribute knowledge, knowledge from pre-trained language models(T5), and corresponding diversity constraint biases into the original data, thereby enhancing its diversity across multiple dimensions.

\textbf{Recommended task fine-tuning.}
During the fine-tuning process for recommendation tasks, we initially integrate the reconstructed project description text with the recommendation task prompt template to fine-tune the LLMs for recommendations. This integration aims to enable the model to comprehend the paradigm of recommendation tasks, and for this purpose, we employ the LoRA fine-tuning method. Subsequently, to enable the LLMs to fully leverage collaborative information, we utilize a pre-trained traditional collaborative model to generate collaborative information vectors for users and items. These vectors are then mapped into the same dimensional vector space as the LLMs using a mapping layer, subsequently replacing the positions of the corresponding ID tokens in the prompt template. To bridge the gap between the collaborative information vectors and the semantics of the LLMs, we focus on fine-tuning the parameters of the mapping layer. This alignment ensures that the collaborative information vectors are congruent with the semantic information of the LLMs, thereby enhancing the model's utilization of traditional collaborative information to improve the accuracy of recommendation tasks. 

Specifically, in fine-tuning the LLMs learning recommendation task paradigm, we take the following form:

$$P_{ui} = LLMs_{Rec}(T_{rec}^{seq} + [d_1^{r}, d_2^{r}, d_3^{r},...,d_n^{r}]),$$

Where \( T_{rec}^{seq} \) denotes the prompt template for sequential recommendation tasks, the sequence \([d_1^{r}, d_2^{r}, d_3^{r}, \ldots, d_n^{r}]\) encapsulates the historical interaction data between user \( u \) and item \( i \). The project description is reconstructed utilizing a trained diversity encoder. In this paper, the description of a project is limited to its title, with a maximum token length of 20. The Large Language Model employed for recommendation purposes, denoted as \( LLMs_{Rec}(\cdot) \), is the LLama2(7B) model, and \( P_{ui} \) represents the predicted outcome of the recommendation model.

Through the fine-tuning process described above, by integrating the project description text with the prompt template, the LLMs has learned the structure of recommendation tasks. Subsequently, we employ a pre-trained traditional model to generate collaborative information vectors for users and items, and we define a mapping layer to align these vectors with the dimensionality of the LLMs. Finally, we replace the positions of the user and item ID tokens in the prompt template with the aligned collaborative information vectors, thereby infusing collaborative information into the LLMs. The specific implementation is as follows: 
$$emb_{u}^{col}, emb_{i}^{col}, P^u_{col} = MAP_{MLP}(Model_{col}(u,i^{tar},[i_1, i_2, i_3,...,i_n])),$$
$$P_{ui} = LLMs_{Rec}(T_{rec}^{seq} + [d_1^{r}, d_2^{r}, d_3^{r},...,d_n^{r}],emb_{u}^{col}, emb_{i}^{col}, P^u_{col}),$$

Here, \( Model_{col}(\cdot) \) represents a pre-trained traditional recommendation model that generates collaborative information vectors \( emb_{u}^{col} \) for user \( u \) and \( emb_{i}^{col} \) for item \( i \); \( MAP_{MLP} (\cdot)\) is the defined mapping layer, which is structurally similar to \( D_{MLP}(\cdot) \), except that its input dimension corresponds to the dimension of the traditional recommendation model, and its output dimension matches the dimension of the LLMs's latent feature vectors. Ultimately, the collaborative information vectors \( emb_{u}^{col} \) and \( emb_{i}^{col} \), along with the prompt template $T_{rec}^{seq} $ and the reconstructed diverse item descriptions $[d_1^{r}, d_2^{r}, d_3^{r},...,d_n^{r}]$, are inputted into the base recommendation model for making recommendations. In this scenario, only the parameters of \( MAP_{MLP} (\cdot)\) are fine-tuned to enable the LLMs to comprehend collaborative information, thereby enhancing the accuracy of recommendation outcomes.

\subsection{Two-stage optimization}

Now, we consider how to optimize the model parameters. First, to obtain diverse item representations, we construct a diversity encoder optimization module based on adversarial generative networks. This module generates diverse item feature vectors, which are then used to fine-tune LLMs, enhancing their robustness. 
Second, to achieve the recommendation task, we utilize the Lora module and mlp mapping layer to assist the LLMs in learning the recommendation task and collaboration information.
By freezing the parameters of the LLMs and focusing on optimizing the Lora module and mlp mapping layer, we accelerate the tuning process of the LLMs. 
One straightforward approach to address this training challenge is to integrate the aforementioned modules into a cohesive end-to-end framework and train them concurrently. However, the ongoing evolution of the diverse encoder modules leads to continuous fluctuations in the reconstructed information of the original project descriptions, potentially undermining the consistency of textual comprehension within the recommendation dataset. To mitigate this issue, we have implemented a two-phase optimization strategy, comprising the \textit{optimizatio of diversity encoder} and the \textit{fine-tuning of recommended LLMs}. 

\subsubsection{Stage 1: Optimization of Diversity Encoder}

In the diversity encoder enhancement phase, we construct the generative adversarial network structure by categorizing the data samples with added attributes into multiple classes and constructing an MLP-based discriminator to classify the embeddings of the data samples. In each training iteration, the generator and discriminator are alternately optimized until convergence.

\textbf{Generator optimization.}
First, for the generator, which is also known as the diversity encoder, the loss function employed is the cross-entropy loss function. This loss function measures the difference between the target ranked sequence and the sequence that has been predicted by the model.:

$$\mathcal{L}_{G_{LLMs_e}} = -\sum_{t=1}^{T} \log P(y_t \mid y_{<t}, \mathbf{x}),$$

where $y_t$ is the $t$th token of the target sequence, $y_{<t}$ is all the tokens before the tth token, $\mathbf{x}$ is the input sequence of the tokens, and $P(y_t \mid y_{<t}, \mathbf{x})$ is the probability predicted by the model for the target word $y_t$ given the input sequence and all previous target words. The model uses this probability distribution to determine the most likely next token in the sequence. To enhance the model’s performance, we introduce a loss function that adds a diversity constraint, aiming to encourage the model to generate more diverse and less repetitive sequences. The total loss function that incorporates the diversity constraint on this basis is as follows:

$$\mathcal{L}_{G} = \mathcal{L}_{G_{LLMs_e}} + \gamma D_{total}\\=-\sum_{t=1}^{T} \log P(y_t \mid y_{<t}, \mathbf{x}) + \gamma(\alpha D_{cos}(x_i \parallel x_i^{\prime}) + \beta D_{JS}(x_i \parallel x_i^{\prime}))),$$

where $\mathcal{L}_{G_{LLMs_{e}}}$ is the loss function of the Large Language Model encoder. The term $D_{total}$ represents the diversity constraint on the data samples, which ensures that the data is diverse enough to be representative of the whole distribution. The parameter $\gamma$ is used to control the strength of this diversity constraint, essentially balancing between the accuracy of the generated samples and their diversity.

The objective of the optimization process for the generator is to minimize the loss function $\mathcal{L}_{G}$. This requires the generator to adjust its parameters in order to generate data samples that not only fit the target distribution well but also meet the diversity constraint imposed by $D_{total}$. Formally, the optimization objective of the generator can be expressed as:

$$\theta_{G} \leftarrow \arg \min _{\theta_{G}} \mathcal{L}_{G},$$

$\theta_{G}$ is the optimized parameter in the generator, representing the parameter that has been fine-tuned during the training process to ensure the best possible performance of the LLMs's encoder.

\textbf{Discriminator optimization.}
For the discriminator $D_{MLP}$, which refers to the multilayer perceptron (MLP) classification model, we employ a loss function of multiclassification. This loss function is crucial for measuring the performance of the model in distinguishing between multiple classes. The computation of this loss function is done in the following manner:

$$\mathcal{L}_{D_{MLP}} = - \sum_{i=1}^{C} y_i \log(\hat{y}_i),$$

In this context, where $y_i$ represents the actual category of the data attribute and $\hat{y}_i$ denotes the predicted category of the data attribute, the objective of the discriminator is to optimize its performance by minimizing the objective function $\\mathcal{L}_{D_{MLP}} $. This optimization process ensures that the discriminator learns to better distinguish between different categories, enhancing its ability to make accurate predictions. The mathematical representation of this objective is given as follows:

$$\theta_{D} \leftarrow \arg \min _{\theta_{D}} \mathcal{L}_{D_{MLP}}$$

$\theta_{D}$ is the parameter to be optimized in the discriminant. It represents the set of weights and biases used by the model to make classifications based on the input features.

\subsubsection{Stage 2: Fine-tuning of Recommended LLMs}

During the fine-tuning phase of the LLMs-based recommendation system, we employ a two-step fine-tuning methodology. Initially, the Large Language Model is fine-tuned using the LoRA technique in conjunction with the reconstructed diverse item descriptions. This process is designed to enable the LLMs to assimilate the nuances of the recommendation task paradigms. 
Then, the mapping layer parameters of mapping collaboration information into the LLMs hidden space are optimized to align the collaboration information vector with the LLMs semantic space so that the LLMs can make full use of the collaboration information.
Specifically, we use binary cross entropy loss (BCELoss) as an objective function for the unity of the two-step recommendation task:

$$\mathcal{L}_{\text{Rec}} = - \frac{1}{N} \sum_{i=1}^{N} \left[ y_i \log (\hat{y}_i) + (1 - y_i) \log (1 - \hat{y}_i) \right],$$

In this context, $y_i$ represents the true label for the $i$-th sample, whereas $\hat{y}_i$ is the predicted probability by the model for the $i$-th sample. The predicted outcome for each sample is expressed through the log-probabilities of the corresponding token in the LLMs's vocabulary. 
The model parameters that need to be optimized at this stage are the parameters of Lora module. Formally, this can be expressed as:

$$\theta_{Rec} \leftarrow \arg \min _{\theta_{D}} \mathcal{L}_{Rec},$$

The parameter $\theta_{Rec}$ represents the model parameters that need to be optimized in the recommendation task. This includes the parameters of mapping layer and the Lora module.

\section{EXPERIMENTS}

In this section, we first introduce the research questions to be answered in the experiment, then present the specific details of the experiment, and finally provide a detailed analysis of the following research questions.

\begin{itemize}
\item [\textbf{RQ1:}] Whether data diversity is significantly enhanced by the GANs-based diversity encoder?
\item [\textbf{RQ2:}] Whether GANPrompt can improve the recommended performance in the case of data diversity?
\item [\textbf{RQ3:}] How do the different modules work in the experiments?
\end{itemize}

\subsection{Experiment Setting}

This section details the datasets used in the experiments, the baseline methods, the evaluation metrics for the comparisons, and the deployment details of the models in this paper.

\textbf{Datesets.}
We conducted experiments on the \textbf{public Amazon dataset} and a \textbf{real large-scale industrial dataset} to evaluate GANPrompt. 
Amazon dataset is a product review dataset that records user review behaviour across multiple domains. The recorded information includes user ID, product ID, user comments text, corresponding ratings and timestamps for the interactions, as well as the product's title and profile description.
The real large-scale industrial dataset is collected from the behavioral logs of the users of the Lenovo online mall, including the records of users' interactions with the products, the time of the interactions, as well as the titles of the products, and the characteristics of various categories.
To evaluate GANPrompt on the Amazon dataset, we selected four sub-datasets: \textbf{Amazon-Books, Amazon-Beauty, Amazon-Toys and Amazon-Sports}. Following common data preprocessing methods\cite{zhang2023recommendation}, we only consider user ratings of $5$ ($R=5$) as users having interactions with the item, i.e., it means that the user prefers the item. Afterwards, the interaction records between users and items are orGANsised chronologically according to timestamps. In addition, we filter out unpopular items and inactive users with less than $5$ interaction records. Note that due to the sparsity of the data in the Amazon-Books dataset, we filter out users and items with fewer than 20 interactions. In addition, we adopt leave-one-out strategy for performance evaluation. Specifically, given a sequence $S = \{i_1, i_2, ..., i_n \}$, we use the last interaction $i_n$ for model testing, the interaction $i_{n-1}$ and the earlier interaction ($\{i_1, i_2, ..., i_{n-2}\}$ ) are used for model training; note that for each data sample, we randomly select 9 items which users have not interacted as negative samples. For the industrial dataset, we use the same data processing strategy. For the sequence length of the interacting data in all datasets, we set the truncation threshold $[5-20]$. Table \ref{tab2} gives the statistics for these datasets.

\begin{table}
    \centering
        \begin{tabular}{cccccc} 
        \hline
        \textbf{Dataset} & \textbf{\#User} & \textbf{\#Item} & \textbf{\#Interaction} & \textbf{\#Sparsity}  \\
        \hline
        Books           & 7763            & 6926            & 238062                & 99.56\% \\ 
        Beauty          & 2267            & 1557            & 19119                 & 99.46\% \\
        Toys            & 1774            & 1390            & 13106                 & 99.47\% \\
        Sports          & 6667            & 4021            & 51550                 & 99.81\% \\
        lenovo          & 4027            & 3075            & 54838                 & 99.56\% \\
        \hline
        \end{tabular}
    \caption{Statistics on preprocessed data sets}
    \label{tab2}
\end{table}

\textbf{Baselines.}
In order to evaluate the recommendation performance of our model, we have used two mainstream recommendation models, including traditional sequence recommendation methods and LLMs-based recommendation methods.

\begin{itemize}
    \item \textbf{Traditional sequence recommendation methods}
    \begin{itemize}
        \item BERT4Rec \cite{sun2019bert4rec}: A Transformer-based sequential recommendation model that bi-directionally encodes a user's sequence of historical behaviours by autoregression to capture complex contextual dependencies to enhance recommendation effectiveness.
        \item SASRec \cite{kang2018self}: It is a sequence recommendation model based on the self-attention mechanism, which achieves efficient personalised recommendation by capturing long-range dependencies in user behaviour sequences.
        \item Caser \cite{tang2018personalized}: Applying horizontal and vertical convolution on a user's historical behavioural sequence to capture sequential patterns and make recommendations.
        \item DIN \cite{zhou2018deep}: Using an attention mechanism to dynamically capture user interests and match them with target items to improve recommendation results.
        \item GRU4Rec \cite{tan2016improved}: Modelling user sequential behaviours and making recommendations using GRUs.
    \end{itemize}
    
    \item \textbf{LLMs-based recommendation methods}
    \begin{itemize}
        \item TallRec \cite{bao2023tallrec}: By structuring recommendation data into instructions and using a lightweight tuning approach to align LLMs for recommendation tasks, to improve the performance of LLMs in the recommendation domain and show strong generalisation capabilities across domains.
        \item CoLLM \cite{zhang2023collm}: Effective integration of collaborative information in recommendation tasks by combining collaborative information with a large-scale language model, utilising collaborative embeddings generated by external conventional collaborative models and mapping them to the input embedding space of the language model.
    \end{itemize}
\end{itemize}

\textbf{Evaluation Metrics.}

In measuring recommendation performance, we use well-accepted evaluation metrics that have been used in previous recommendation methods. These metrics include Area Under the Curve ($AUC$), Normalised Discounted Cumulative Gain ($NDCG@k$), Hit Rate ($HR@k$), and Mean Reciprocal Ranking ($MRR@k$), where $k=[1, 3, 5]$ is varied.
The $AUC$ measures the model's ability to distinguish between positive and negative samples and is commonly used for recommender system evaluation in binary classification tasks.
The $HR@k$ is used to evaluate whether the recommender system contains items that are actually of interest to the user in the first $k$ recommended items, reflecting the coverage and accuracy of the recommendation results.
The $NDCG@k$ comprehensively evaluates the relevance and sorting quality of the recommender system by considering the rank order of items in the recommended list, with higher values indicating that the system not only accurately recommends items, but also efficiently sorts relevant items to a higher position.
$MRR@k$ evaluates the accuracy of the recommender system by calculating the reciprocal of the position of the first correctly recommended item among the first k recommendations, and higher MRR values indicate that the recommender system is able to recommend the items of interest to the user at the top more frequently. Both the $NDCG$ and $MRR$ metrics focus more on the quality of the recommendation ranking.
All metrics range from $0$ to $1$, with higher values indicating better results.

\textbf{Implementation Details.}
All experiments were conducted in a Python 3.9 environment using the PyTorch framework and CUDA version 12.3, on NVIDIA's A800 80G GPUs. Specifically, for the attribute generation segment, we utilized BERT as the attribute generator in a masked tokens prediction task, selecting the top five attributes with the highest prediction frequency from each dataset as additional features. 
During the diversity encoder training phase, GANPrompt employed the T5 model’s encoder as the generator in the framework of Generative Adversarial Network, with a three-layer Multilayer Perceptron (MLP) serving as the discriminator; the batch size was set to 256, and both the encoder and discriminator were trained with a learning rate of 1e-5 using the Adam optimizer. For the sequential recommendation task, we used the pre-trained llama2-7B language model as the foundational LLMs for recommendations. The training batch size was 64 with a learning rate of 2e-4, using the AdamW optimizer.

\subsection{Diversity Encoder Validity Experiment}

\begin{figure}[hbp]
    \centering
    \begin{subfigure}[b]{0.4\textwidth}
        \centering
        \includegraphics[width=0.9\textwidth]{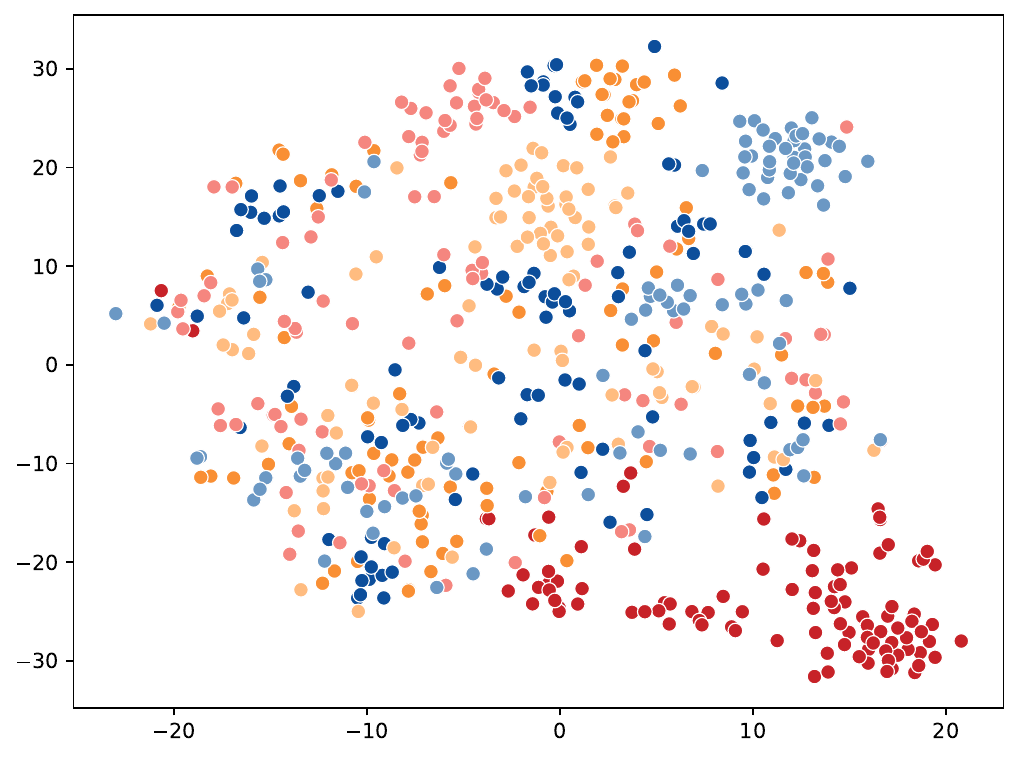}
        \caption{\footnotesize Original titles (Amazon-Books).}
        \label{fig 3.1}
    \end{subfigure}
    \begin{subfigure}[b]{0.4\textwidth}
        \centering
        \includegraphics[width=0.9\textwidth]{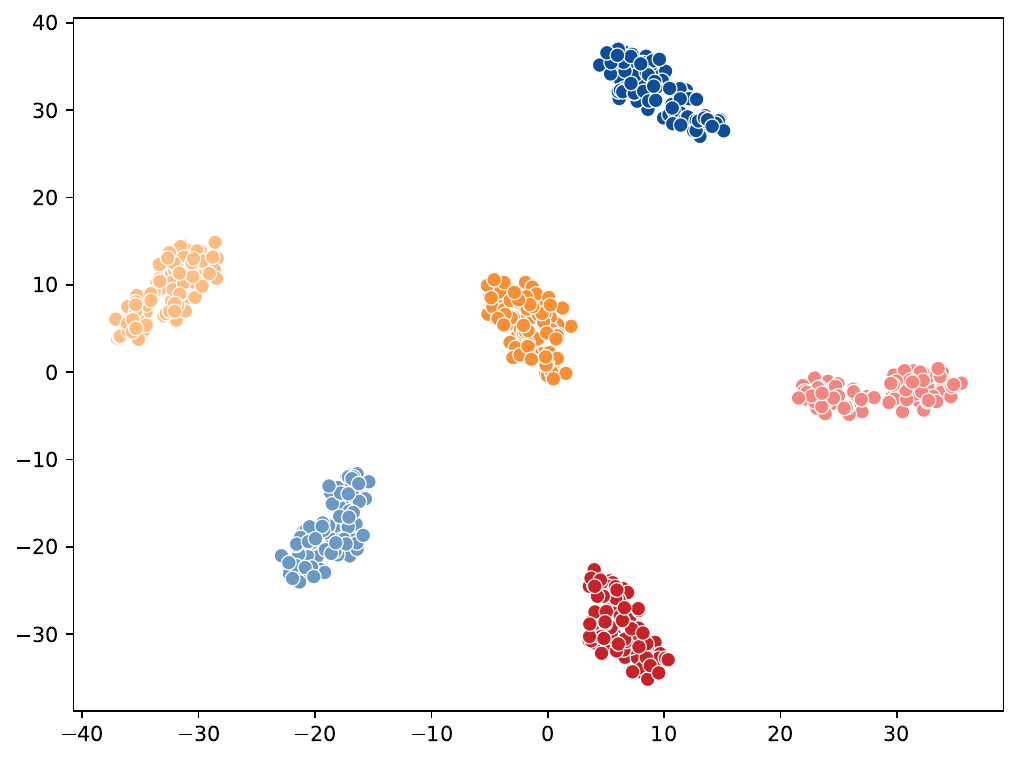}
        \caption{\footnotesize Diversity-enhanced titles (Amazon-Books).}
        \label{fig 3.2}
    \end{subfigure}

    \begin{subfigure}[b]{0.4\textwidth}
        \centering
        \includegraphics[width=0.9\textwidth]{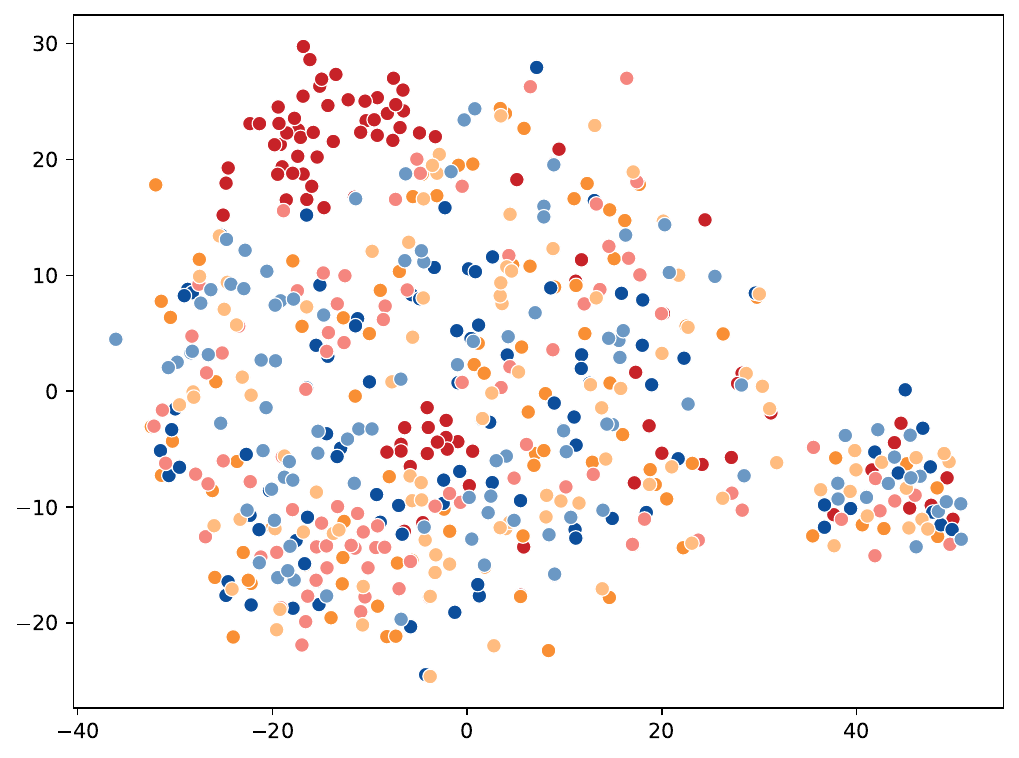}
        \caption{\footnotesize Original titles (Amazon-Beauty).}
        \label{fig 3.1}
    \end{subfigure}
    \begin{subfigure}[b]{0.4\textwidth}
        \centering
        \includegraphics[width=0.9\textwidth]{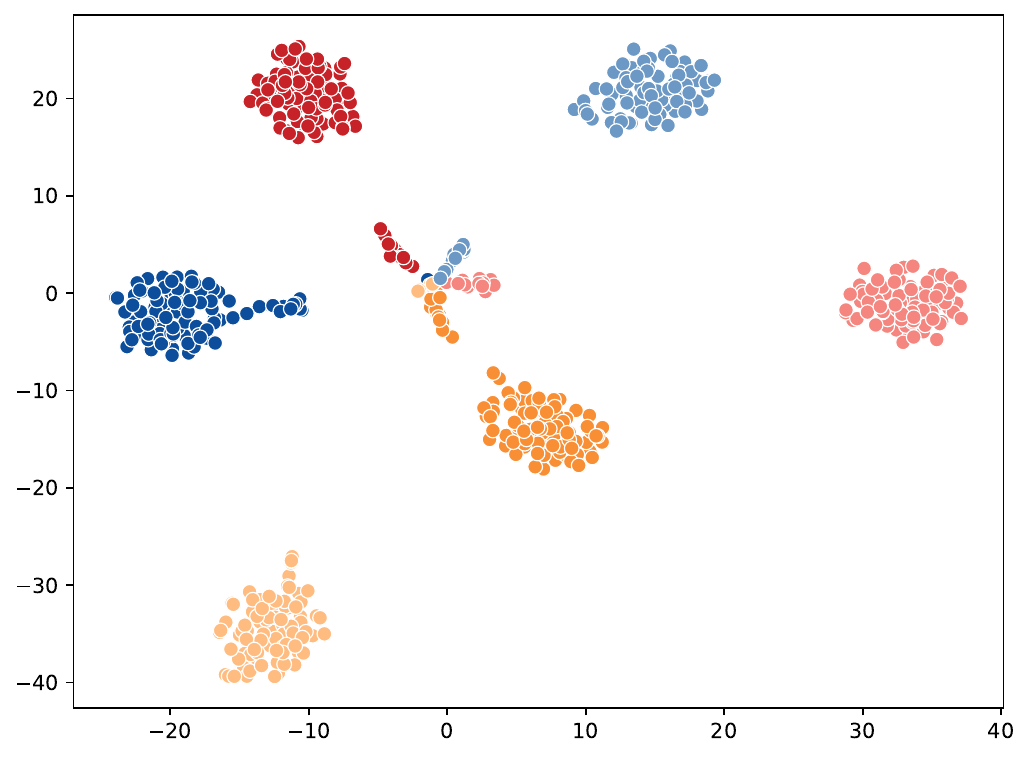}
        \caption{\footnotesize Diversity-enhanced titles (Amazon-Beauty).}
        \label{fig 3.2}
    \end{subfigure}

    \begin{subfigure}[b]{0.4\textwidth}
        \centering
        \includegraphics[width=0.9\textwidth]{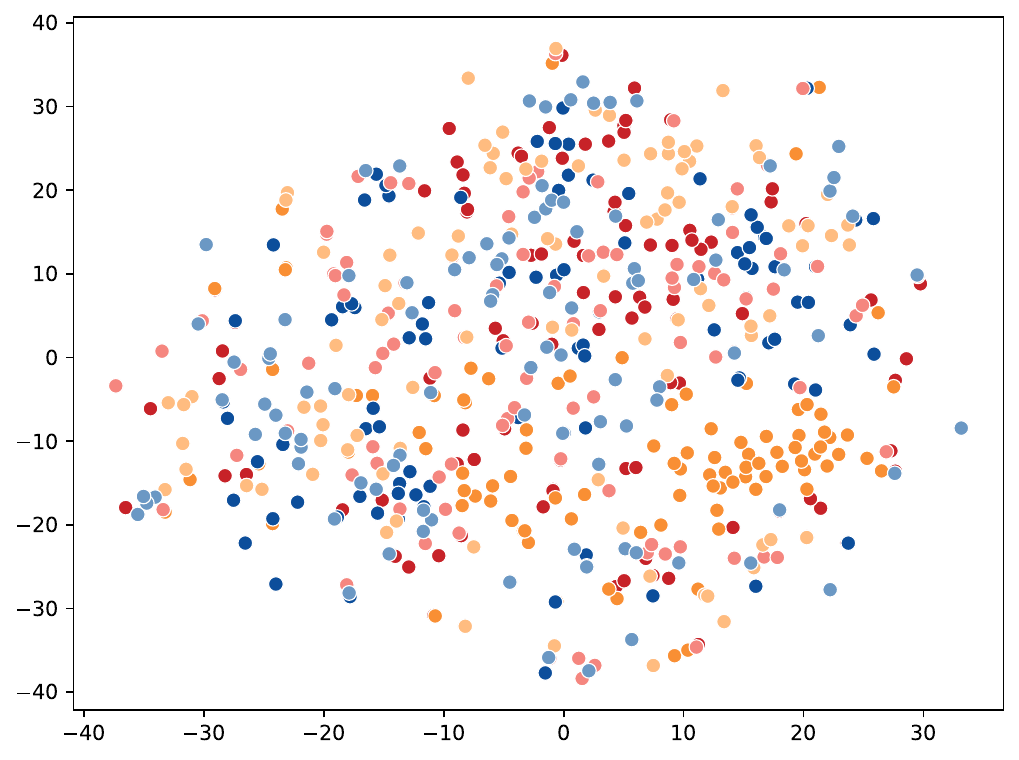}
        \caption{\footnotesize Original titles (Amazon-Toys).}
        \label{fig 3.1}
    \end{subfigure}
    \begin{subfigure}[b]{0.4\textwidth}
        \centering
        \includegraphics[width=0.9\textwidth]{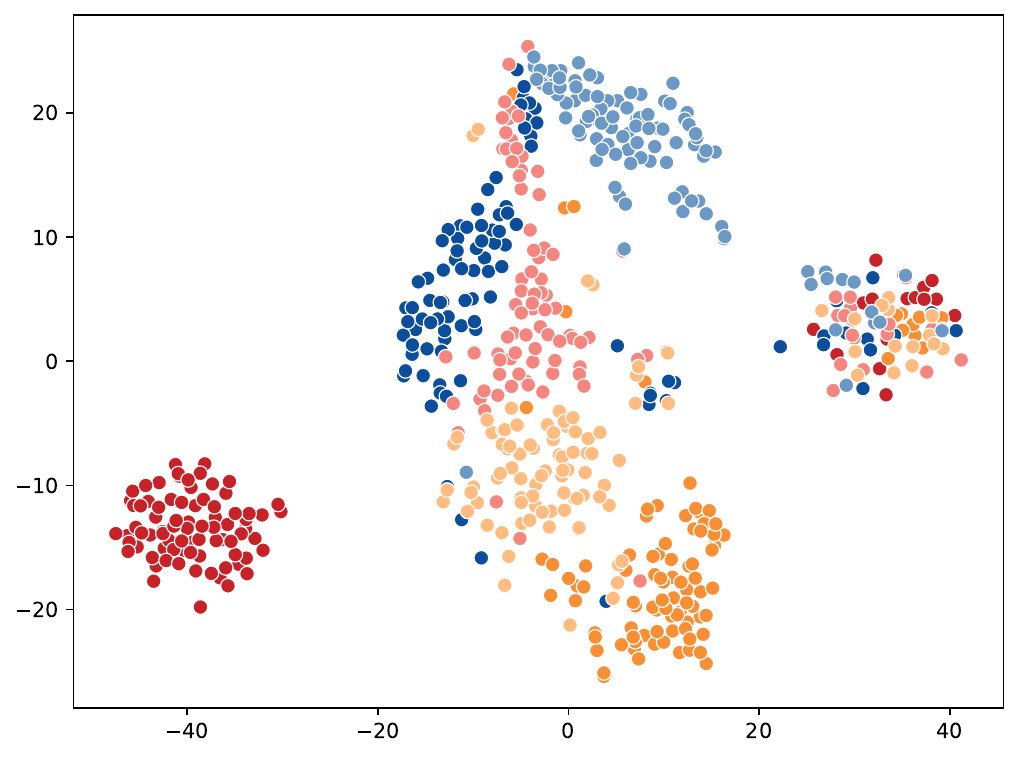}
        \caption{\footnotesize Diversity-enhanced titles (Amazon-Toys).}
        \label{fig 3.2}
    \end{subfigure}

    \begin{subfigure}[b]{0.4\textwidth}
        \centering
        \includegraphics[width=0.9\textwidth]{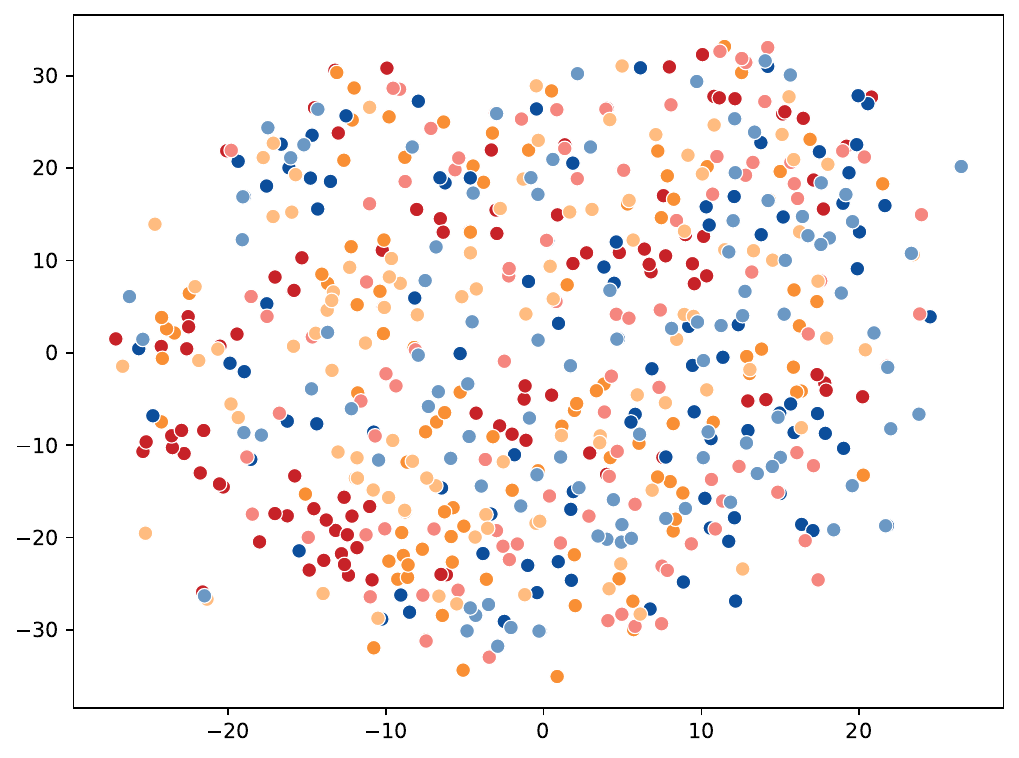}
        \caption{\footnotesize Original titles (Amazon-Sports).}
        \label{fig 3.1}
    \end{subfigure}
    \begin{subfigure}[b]{0.4\textwidth}
        \centering
        \includegraphics[width=0.9\textwidth]{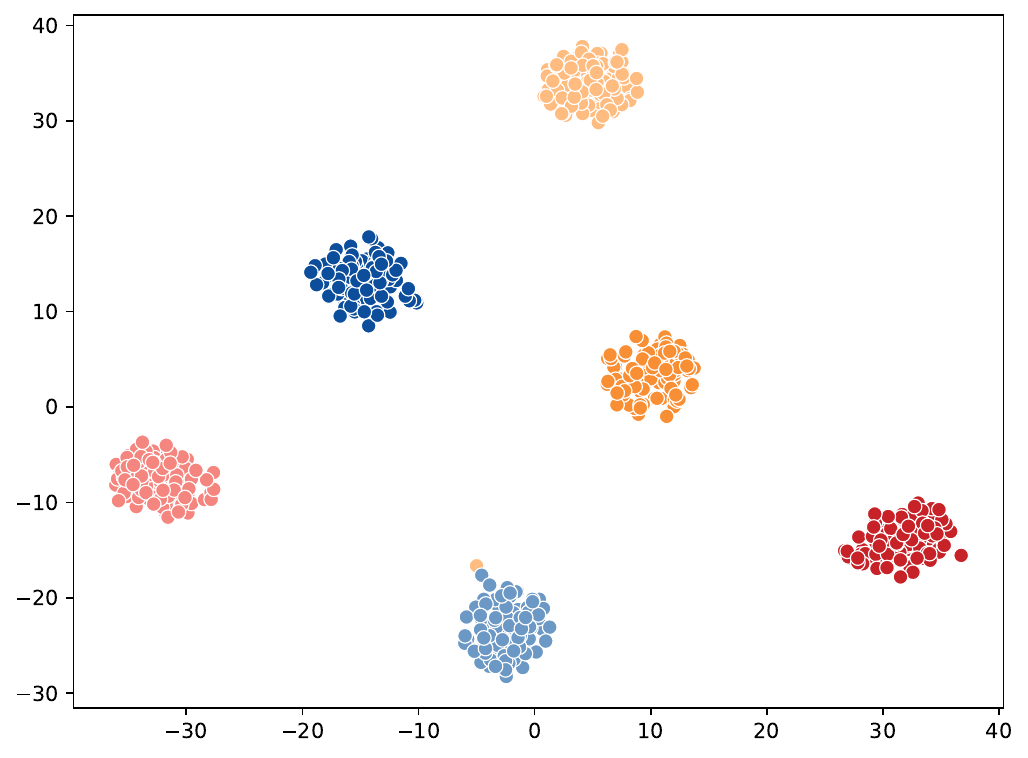}
        \caption{\footnotesize Diversity-enhanced titles (Amazon-Sports).}
        \label{fig 3.2}
    \end{subfigure}
    
    \caption{Visualisation of item title data from four Amazon datasets before and after diversity encoder enhancement, including attribute data after the addition of five attribute features.}
    \label{Div vis}
\end{figure}

In order to verify whether the diversity encoder, which has been adapted by applying the GANs framework, can effectively differentiate data samples and enhance the diversity among data samples and answer the \textbf{RQ1}, we conducted detailed experimental analyses. 



\textbf{Visualisation of sample distribution}

In order to further validate the effectiveness of the diversity encoder, we conducted further visual presentation experiments. Specifically, we performed T-SNE visualisations on four amazon datasets on both the original data and the data embeddings enhanced after diversity. Specifically, , we used the item titles in each dataset as the original inputs, and then combined the first 5 attribute data in each dataset to construct a composite dataset containing 6 attribute dimensions for further visual presentation and analysis.

The visualisation results are shown in Figure \ref{Div vis}. From the figure, it can be observed that the encoder of the pre-trained language model T5 has difficulty in distinguishing datasets with different attributes before training, and all the sample points are almost clustered in the same data distribution region. However, after the diversity constraint adjustment by introducing the GANs framework and diversity encoder, the situation changes significantly. The datasets with different attributes are effectively differentiated in the feature space, and the data points of each collection start to scatter over a wider area.

This obvious differentiation effect indicates that the diversity-adjusted encoder can effectively expand the distance between data with different attributes and change their distribution in the feature space. This not only illustrates the effectiveness of the diversity encoder under the GANs framework in improving data diversity, but also provides richer and more discriminative feature representations for the subsequent sequential recommendation task, enhancing the advantage of the subsequent model in dealing with complex and multi-attribute recommendation system data.

\subsection{Recommended Performance Main Experiment Results}

\begin{table}[]
\begin{tabular}{c|c|cccccccc}
\hline
\textbf{Dataset} & \textbf{Model} & \textbf{AUC} & \textbf{HR@1} & \textbf{HR@3} & \textbf{HR@5} & \textbf{NDCG@3} & \textbf{NDCG@5} & \textbf{MRR@3} & \textbf{MRR@5} \\
\midrule
\multirow{9}{*}{\textbf{Books}}& BERT4Rec & 0.528228 & 0.049581 & 0.205538 & 0.258274 & 0.141985 & 0.163961 & 0.119929 & 0.132273 \\
& SASRec & 0.549223 & 0.124662 & 0.379459 & 0.549388 & 0.272429 & 0.342950 & 0.235523 & 0.274966 \\ 
& Caser & 0.601618 & 0.212492 & 0.465551 & 0.669994 & 0.356996 & 0.440733 & 0.319725 & 0.365917 \\
& DIN & \underline{0.852737} & \underline{0.477527} & \underline{0.841404} & \underline{0.939536} & \underline{0.691486} & \underline{0.732177} & \underline{0.639579} & \underline{0.662319} \\
& GRU4Rec & 0.840848 & 0.443400 & 0.822537 & 0.937090 & 0.665359 & 0.712959 & 0.611011 & 0.637669 \\  \cline{2-10} 
& TALLRec & 0.710388 & 0.287122 & 0.584353 & 0.763619 & 0.458121 & 0.531853 & 0.414692 & 0.455545 \\
& CoLLM & 0.830111 & 0.445654 & 0.800064 & 0.924984 & 0.652198 & 0.703941 & 0.601138 & 0.630020 \\ \cline{2-10} 
& \textbf{\textbf{PromptGANs}} & \textbf{0.862959} & \textbf{0.520927} & \textbf{0.857244} & \textbf{0.940502} & \textbf{0.718307} & \textbf{0.752863} & \textbf{0.670230} & \textbf{0.689560} \\
& \textbf{Impr. (\%)} & 1.20\% ↑ & 9.09\% ↑ & 1.09\% ↑ & 0.10\% ↑ & 3.85\% ↑ & 2.83\% ↑ & 4.80\% ↑ & 4.11\% ↑ \\
\midrule
\multirow{9}{*}{\textbf{Beauty}}& BERT4Rec & 0.562237 & 0.160504 & 0.418487 & 0.621849 & 0.308420 & 0.391878 & 0.270588 & 0.316723 \\
& SASRec & 0.587237 & 0.189916 & 0.435294 & 0.650420 & 0.330099 & 0.418808 & 0.293978 & 0.343263 \\ 
& Caser & 0.649685 & 0.242017 & 0.542017 & 0.731092 & 0.414132 & 0.491732 & 0.370168 & 0.413067 \\
& DIN & \underline{0.822847} & \underline{0.505042} & \underline{0.781513} & \underline{0.895798} & \underline{0.666493} & \underline{0.713577} & \underline{0.626751} & \underline{0.652885} \\
& GRU4Rec & 0.803571 & 0.455462 & 0.760504 & 0.883193 & 0.631749 & 0.682526 & 0.587395 & 0.615714 \\ \cline{2-10} 
& TALLRec & 0.705695 & 0.316807 & 0.576471 & 0.748739 & 0.466553 & 0.537357 & 0.428711 & 0.467913 \\
& CoLLM & 0.798004 & 0.457983 & 0.742016 & 0.873949 & 0.623765 & 0.678449 & 0.582913 & 0.613459 \\\cline{2-10} 
&  \textbf{PromptGANs} & \textbf{0.838445} & \textbf{0.522689} & \textbf{0.800000} & \textbf{0.906723} & \textbf{0.687861} & \textbf{0.731467} & \textbf{0.648880} & \textbf{0.672871} \\
& \textbf{Impr. (\%)} & 1.34\% ↑ & 3.49\% ↑ & 2.37\% ↑ & 1.22\% ↑ & 3.21\% ↑ & 2.51\% ↑ & 3.53\% ↑ & 3.06\% ↑ \\
\midrule
\multirow{9}{*}{\textbf{Toys}}& BERT4Rec & 0.540840 & 0.157407 & 0.384921 & 0.587302 & 0.289002 & 0.371758 & 0.255952 & 0.301521 \\
& SASRec & 0.567295 & 0.158730 & 0.404762 & 0.588624 & 0.301489 & 0.378066 & 0.265873 & 0.308862 \\ 
& Caser & 0.592097 & 0.179894 & 0.456349 & 0.664021 & 0.336999 & 0.422323 & 0.296076 & 0.343298 \\
& DIN & \underline{0.747685} & 0.353175 & \underline{0.677249} & \underline{0.838624} & 0.539977 & \underline{0.605710} & 0.492725 & 0.528770 \\
& GRU4Rec & 0.720155 & 0.318783 & 0.645503 & 0.808201 & 0.505870 & 0.572578 & 0.457892 & 0.494731 \\  \cline{2-10} 
& TALLRec & 0.632863 & 0.244709 & 0.488095 & 0.662698 & 0.386145 & 0.457749 & 0.350970 & 0.390520 \\
& CoLLM & 0.741402 & \underline{0.355079} & 0.669312 & 0.822751 & \underline{0.541096} & 0.604396 & \underline{0.496914} & \underline{0.532099} \\ \cline{2-10} 
& \textbf{PromptGANs} & \textbf{0.763062} & \textbf{0.363757} & \textbf{0.699735} & \textbf{0.850529} & \textbf{0.559629} & \textbf{0.621616} & \textbf{0.511243} & \textbf{0.545569} \\
& \textbf{Impr. (\%)} & 2.06\% ↑ & 2.44\% ↑ & 3.32\% ↑ & 1.42\% ↑ & 3.43\% ↑ & 2.63\% ↑ & 2.88\% ↑ & 2.53\% ↑ \\
\midrule
\multirow{9}{*}{\textbf{Sports}}& BERT4Rec & 0.545437 & 0.102882 & 0.351410 & 0.467927 & 0.248731 & 0.297228 & 0.213201 & 0.240409 \\
&SASRec & 0.518051 & 0.104431 & 0.341804 & 0.559343 & 0.240361 & 0.330533 & 0.205506 & 0.255877 \\ 
&Caser & 0.574295 & 0.175395 & 0.425782 & 0.646731 & 0.317832 & 0.407856 & 0.280808 & 0.330188 \\
&DIN & \underline{0.680411} & 0.251317 & \underline{0.578246} & 0.768826 & 0.440059 & 0.518334 & 0.392470 & 0.435776 \\
& GRU4Rec & 0.683685 & \underline{0.266192} & 0.574837 & \underline{0.779362} & \underline{0.443398} & \underline{0.527394} & \underline{0.398203} & \underline{0.444670} \\ \cline{2-10} 
& TALLRec & 0.616982 & 0.183452 & 0.455222 & 0.649830 & 0.337432 & 0.417457 & 0.297077 & 0.341406 \\
&CoLLM & 0.642973 & 0.235203 & 0.528355 & 0.731329 & 0.401984 & 0.485299 & 0.358641 & 0.404705 \\ \cline{2-10} 
& \textbf{PromptGANs} & \textbf{0.692516} & \textbf{0.269600} & \textbf{0.591261} & \textbf{0.784940} & \textbf{0.456276} & \textbf{0.535764} & \textbf{0.40972}0 & \textbf{0.453662} \\
& \textbf{Impr. (\%)} & 1.78\% ↑ & 1.01\% ↑ & 2.25\% ↑ & 0.72\% ↑ & 2.90\% ↑ & 1.59\% ↑ & 2.89\% ↑ & 2.02\% ↑ \\
\bottomrule
\end{tabular}
\caption{Performance metrics on the Amazon dataset. Maximum values are bolded, second-best values are underlined. Impr. denotes the improvement ratio.}
\label{tab3}
\end{table}

\begin{table}[]
\begin{tabular}{c|c|cccccccc}
\hline
\textbf{Dataset} & \textbf{Model} & \textbf{AUC} & \textbf{HR@1} & \textbf{HR@3} & \textbf{HR@5} & \textbf{NDCG@3} & \textbf{NDCG@5} & \textbf{MRR@3} & \textbf{MRR@5} \\
\midrule
\multirow{7}{*}{\textbf{Industry}}& BERT4Rec & 0.568320 & 0.168364 & 0.421157 & 0.620810 & 0.313098 & 0.395721 & 0.275971 & 0.322047 \\
& SASRec & 0.527595 & 0.119940 & 0.356345 & 0.581078 & 0.255732 & 0.348384 & 0.221132 & 0.272597 \\ 
& Caser & 0.670862 & 0.253042 & 0.560715 & 0.757636 & 0.429345 & 0.510237 & 0.384198 & 0.428959 \\
& DIN & \underline{0.771961} & \underline{0.357090} & \underline{0.711696} & \underline{0.874348} & \underline{0.562582} & \underline{0.629639} & \underline{0.511175} & \underline{0.548423} \\
& GRU4Rec & 0.752002 & 0.321828 & 0.682146 & 0.854482 & 0.530240 & 0.601208 & 0.477899 & 0.517271 \\  \cline{2-10} 
& TALLRec & 0.625528 & 0.195679 & 0.477527 & 0.657561 & 0.355753 & 0.429459 & 0.314005 & 0.354644 \\
& CoLLM & 0.773373 & 0.353365 & 0.711696 & 0.876335 & 0.560687 & 0.628556 & 0.508650 & 0.546346 \\ \cline{2-10} 
& \textbf{PromptGANs} & \textbf{0.781087} & \textbf{0.374969} & \textbf{0.721132} & \textbf{0.884778} & \textbf{0.575654} & \textbf{0.642889} & \textbf{0.525495} & \textbf{0.562706} \\
& \textbf{Impr. (\%)} & 1.00\% ↑ & 5.01\% ↑ & 1.33\% ↑ & 0.96\% ↑ & 2.32\% ↑ & 2.10\% ↑ & 2.80\% ↑ & 2.60\% ↑ \\
\midrule
\end{tabular}
\caption{Performance metrics on the industrial dataset. Maximum values are bolded, second-best values are underlined. Impr. denotes the improvement ratio.}
\label{tab4}
\end{table}

In order to answer \textbf{RQ2}, the performance of GANPrompt and all baseline models on the five datasets is shown in table \ref{tab3} and \ref{tab4}, with the best results are indicated in bold and second-best results are underline.

The results of the experiment show some interesting elements. Firstly, looking at the five datasets as a whole, the effectiveness of TALLRec, which is LLMs fine-tuned using only textual corpus information, suggests that the LLMs are able to understand the textual form of the recommendation task and learn the corresponding task paradigm. In contrast, traditional deep neural network models using collaborative signals are not inferior or even can take better results, e.g., DIN, GRU4Rec, suggesting that collaborative signals play a crucial role in recommendation tasks and that textual corpus information alone is not sufficient for adequate recommendation. On this basis, CoLLM combines textual knowledge and collaborative information to collaboratively fine-tune the LLMs and achieve efficient results. 
Finally, by fine-tuning robustness and accuracy on diversity-enhanced datasets as well as using expert-guided collaborative signalling, PromptGANs achieves the best results on all metrics.

In terms of each dataset, the Books and Beauty datasets have lower data sparsity and richer information about the interactions between users and items; in addition, their text corpus information is also richer and more relevant and logical relative to the datasets of Toys, Sports and Industry, which enables the models to learn more information and thus achieve higher recommendation results. 
Specifically, using the book dataset as an example, PromptGANs improves 1.20\% relative to the second-best traditional collaborative model DIN and 3.96\% relative to CoLLM, a recommendation method based on LLMs, on the auc metrics.
Specifically, for the AUC metric, taking the book dataset as an example, PromptGANs improves 1.20\% on the auc metrics relative to the suboptimal traditional collaborative model DIN, and 3.96\% relative to CoLLM, the LLMs-based recommendation method that co-operatively utilises collaborative and semantic information.
This is due to the fact that traditional sequential recommendation methods, which focus on more streamlined structure and learning tasks, excel at capturing the changing interests of users in recommendation scenarios through collaborative information between user items, and thus excel in personalised recommendation accuracy. 
CoLLM utilises the semantic understanding and generation capabilities of LLM and complements them with traditional collaborative information to achieve excellent positive and negative sample differentiation.
PromptGANs, on the other hand, fine-tunes on the diversity encoder-enhanced instruction dataset to enhance the robustness of LLMs; and uses expert-guided tokens to enhance the understanding of LLMs on the knowledge of the traditional collaborative model, thus achieving further performance improvement.
On the HR@k, NDCG@k and MRR@k metrics, PromptGANs improves its performance by 2.55\% on average relative to baselines, and even by 9.09\% on the $HR@1$ metric on the books dataset.
This is because HR@k, NDCG@k, and MRR@k are more focused on measuring the ranking ability of the model front-end, i.e., the exact position of the correct sample in the recommendation list, relative to AUC, which focuses on distinguishing between positive and negative samples.
LLMs-based recommendation methods use semantic information to decipher the correlation between items and user interests, thus accurately placing the more relevant items at the top of the recommendation list, resulting in a stronger ranking capability. CoLLM and PromptGANs, on the other hand, combine the semantic understanding of LLMs and the collaborative information of traditional collaborative models, and take into account both positive and negative sample classification and sample rating ranking to achieve superior recommendation performance.
In addition, PromptGANs is fine-tuned on the basis of diversity-enhanced data and receives collaborative knowledge from traditional models in a form that is more closely aligned to the semantic space, resulting in greater resistance to perturbation and fuller exploitation of traditional model knowledge, with greater stability and knowledge exploitation leading to superior performance.

\subsection{Ablation Study}
In order to answer \textbf{RQ3} and test the validity of the different modules in the model, we performed ablation experiments by shielding the different modules.

\begin{figure}[h]
  \centering
  \includegraphics[width=0.6\linewidth]{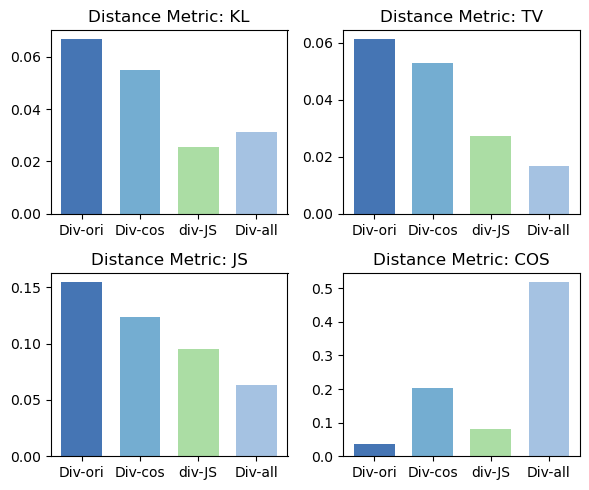}
  \caption{The results of diversity encoder ablation study on the Beauty dataset.}
  \label{ablation div1}
\end{figure}

\begin{figure}[h]
  \centering
  \includegraphics[width=0.6\linewidth]{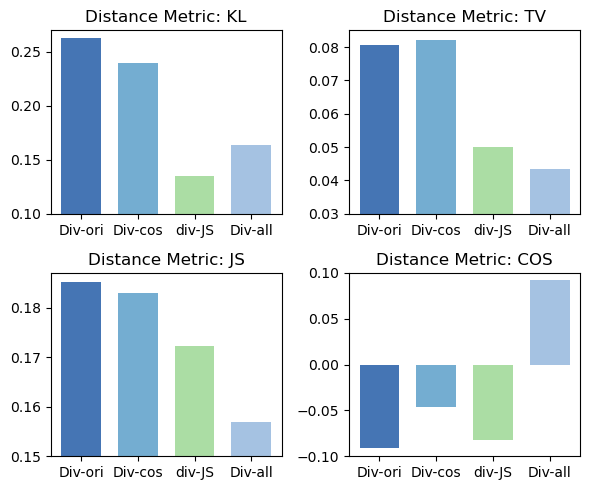}
  \caption{The results of diversity encoder ablation study on the Books dataset.}
  \label{ablation div2}
\end{figure}

Firstly, to evaluate the effectiveness of the proposed diversity constraints, three model variants were constructed for the training of the Diversity Encoder:
\begin{itemize}
    \item \textbf{Div-ori:} Our model with all diversity constraint (cosine similarity and the Jensen-Shannon divergence).
    \item \textbf{Div-cos:} The cosine similarity is removed from the diversity constraint.
    \item \textbf{Div-JS:} The Jensen-Shannon divergence is removed from the diversity constraint.
    \item \textbf{Div-all:} Both the cosine similarity and the Jensen-Shannon divergence are removed from the diversity constraint.
\end{itemize}
The specific experimental results of Amazon-Beauty and Amazon-Books datasets are shown in Figure \ref{ablation div1} and Figure 
 \ref{ablation div2}.

Based on the theory that diversity and distance among sample data are positively correlated, we have adopted four distance metrics to measure the distribution of diversity among samples: Kullback-Leibler (KL) divergence, total variation (TV), Jensen-Shannon (JS) divergence, and cosine similarity. KL divergence, total variation, and JS divergence are positively correlated with distance, where higher values indicate greater distance, implying stronger diversity among samples. In contrast, the cosine similarity constraint is negatively correlated with distance, with lower values indicating greater differences.
From the charts, it is evident that the worst results were obtained from the model without any diversity constraints, Div-all, suggesting that both cosine similarity and JS divergence constraints can effectively optimize the diversity among samples. Overall, Div-cos and Div-JS performed more excellently, achieving second-best results in all but the cosine metric, indicating that the angle-related distance constraint of cosine similarity is slightly weaker than the information-related distance constraint of JS divergence.
The model with the diversity constraints proposed in this paper, Div-ori, achieved the best results in all metrics, demonstrating that the synergistic effect of cosine similarity and JS divergence can expand the distances between data samples from multiple dimensions, thereby effectively enhancing data diversity.

\begin{figure}[h]
  \centering
  \includegraphics[width=0.8\linewidth]{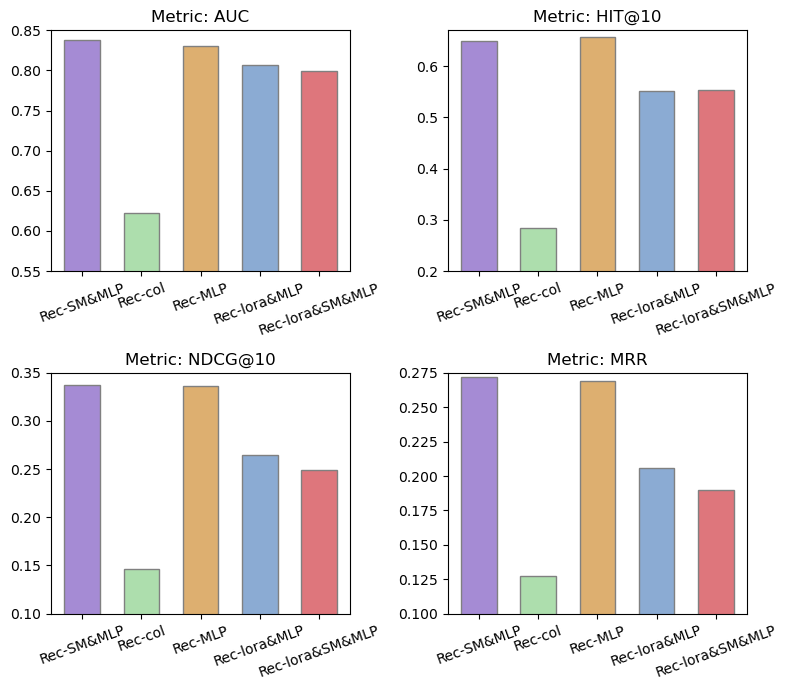}
  \caption{The results of diversity encoder ablation study on the Books dataset.}
  \label{ablation rec1}
\end{figure}

Secondly, to assess the effectiveness of collaborative information and two-stage training in the recommendation task, we proposed two model variants.
\begin{itemize}
  \item \textbf{Rec-ori:} Our model with collaborative signal and only train the parameters of MLP mapping layers in the phase of LLMs recommendation task fine-tuning.
  \item \textbf{Rec-col:} The collaborative signal in the recommendation task is 
  removed.
  \item \textbf{Rec-lora:} The LLMs recommendation task fine-tuning phase simultaneously fine-tunes the parameters of the LoRA and MLP mapping layers.
  \item \textbf{Rec-sm:} The LLMs recommendation task fine-tuning phase simultaneously fine-tunes the parameters of the traditional collaborative model and MLP mapping layer.
  \item \textbf{Rec-lora\&sm:} The LLMs recommendation task fine-tuning phase simultaneously fine-tunes the parameters of the LoRA, the traditional collaborative model and MLP mapping layers.
\end{itemize}
The specific experimental results are shown in Figure \ref{ablation rec1}.

\section{CONCLUSIONS}
In this paper, we present an innovative framework called GANPrompt, which enhances the robustness of LLMs in recommendation systems using GANs. Our research focuses on enhancing the model's adaptability and stability to different prompts through diverse prompt generation. Specifically, GANPrompt first trains a multidimensional prompt generator through GANs generation techniques, which is capable of generating highly diverse prompts based on user behavioural data. These diverse prompts are then used to train the LLMs to improve its performance when faced with unseen prompts. Furthermore, to ensure that the generated prompts are both highly diverse and relevant, we introduce a diversity constraint mechanism based on mathematical theory to optimise prompt generation and ensure that they semantically cover a wide range of user intentions.

Through extensive experiments on multiple datasets, we demonstrate the effectiveness of the proposed framework, especially in improving the adaptability and robustness of recommendation systems in complex and dynamic environments. Experimental results show that GANPrompt provides significant improvements in accuracy and robustness compared to existing state-of-the-art approaches. These results not only demonstrate the effectiveness of diversity encoders under the GANs framework in enhancing the diversity of data, but also provide new technical paths and strong experimental evidence for processing complex and multi-attribute recommendation system data. Advantages when processing complex and multi-attribute recommendation system data.


\bibliographystyle{ACM-Reference-Format}
\bibliography{ref}

\end{document}